\documentclass[useAMS,usenatbib,usegraphicx]{mn2e}
\usepackage{amsmath,amssymb,amsfonts}
\usepackage{times}
\usepackage{url}
\usepackage{pstricks}

\newcommand{\dif}[2]{\frac{\partial #1}{\partial #2}}

\newcommand{\includegraphflex}[2][scale=1,clip]{\includegraphics[#1]{./images_cv_eps/#2.eps}}

\title[Convective viscosity in accretion discs]%
{Turbulent viscosity by convection in accretion discs --\\ a self-consistent approach}
\author[D. Heinzeller, W.J. Duschl and S. Mineshige]{D. Heinzeller$^{1,2}$\thanks{E-mail:
dominikus@kusastro.kyoto-u.ac.jp}, W.J. Duschl$^{2,3}$ and S. Mineshige$^{1}$\\
$^{1}$Department of Astronomy, Graduate School of Science, Kitashirakawa-Oiwakecho, Sakyu-ku, Kyoto 606-8502, Japan\\
$^{2}$Institut f{\"u}r Theoretische Physik und Astrophysik, Leibnizstra{\ss}e 15, 24118 Kiel, Germany\\
$^{3}$Steward Observatory, The University of Arizona, 933 North Cherry Avenue, Tucson, AZ 85721, USA}
\begin{document}%
% Accepted 2009 April 29. Received 2009 April 26; in original form 2009 March 23
\date{Accepted 2009 April 29. Received 2009 April 26; in original form 2009 March 23}%
\pagerange{\pageref{firstpage}--\pageref{lastpage}} \pubyear{2009}%
\maketitle%
\label{firstpage}%
\begin{abstract}%
The source of viscosity in astrophysical accretion flows is still a hotly debated issue. We investigate the contribution of convective turbulence to the total viscosity
in a self-consistent approach, where the strength of convection is determined from the vertical disc structure itself. Additional sources of viscosity are parametrized
by a $\beta$-viscosity prescription, which also allows an investigation of self-gravitating effects. In the context of {accretion discs around stellar mass and intermediate mass black holes},
we conclude that convection alone cannot account for the total viscosity in the disc, but significantly adds to it. For accretion rates up to $10\%$ of the Eddington rate,
we find that differential rotation provides a sufficiently large underlying viscosity. For higher accretion rates, further support is needed in the inner disc region, which
can be provided by an MRI-induced viscosity. We briefly discuss the interplay of MRI, convection and differential rotation.
We conduct a detailed parameter study of the effects of central masses and accretion rates on the disc models {and find that the threshold
value} of the supporting viscosity is determined mostly by the Eddington ratio with only little influence from the central black hole mass.
\end{abstract}%
\section{Introduction}\label{sec_introduction}
Modern theoretical modeling of accretion discs dates back to the year 1948, when Weizs\"{a}cker published his article about the rotation of cosmic
gas \citep{weizsaecker_1948}. A key ingredient to describe the accretion process is the origin of the viscosity, which causes friction
in the disc and an inward motion of the material. First observations of accretion timescales of discs in cataclysmic variables 
{invalidated} theoretical expectations of molecular viscosity being the driving force \citep[see, e.\,g.,][]{prendergast_1968,pringle_1972}. Adversely,
they revealed a discrepancy of many orders of magnitude between the numbers measured in the lab and those needed to account for the observations.
{Soon thereafter, \citet{shakura_1973} proposed the $\alpha$-viscosity parametrization, by which most observations could be reproduced satisfactorily.
Nevertheless, the Shakura-Sunyaev viscosity remains a purely empirical description and is limited to thin discs with negligible disc masses (i.\,e.,
non-selfgravitating discs).} Among the physical theories, the most promising ones are:

\medskip%
\textit{Differential rotation.} An obvious candidate for the turbulence in nearly Keplerian rotating discs is differential rotation.
From early laboratory experiments on rotating Couette-Taylor flows \citep{wendt_1933,taylor_1936}, this possibility was first ruled out.
However, in recent re-investigations, \citet{richard_1999} and \citet{richard_2001} concluded that differential rotation can give rise to turbulence,
despite published arguments. At the same time, \citet*{duschl_1998,duschl_2000} formulated the $\beta$-viscosity description. Although being a
parametrization like its ancestor, it can actually be related to the process of differential rotation. Contrary to the $\alpha$-prescription,
the $\beta$-viscosity accounts properly for the selfgravity of the disc. At the same time, it includes the $\alpha$-viscosity in the case
of a shock dissipation limited, non-selfgravitating disc. Combining the laboratory measurements with the formulation of the $\beta$-viscosity
leads to a value of $\beta\approx10^{-5}$ which can be provided by differential rotation.\footnote{As a rule of thumb, $\beta \approx \alpha^2 \ldots \alpha$ \citep{duschl_2000};
the corresponding $\alpha$ parameter therefore lies between $10^{-5}$ and $10^{-3}$.}

\textit{Convection.} In order to account for the transport of the energy released by the accretion process, convection is considered to support or even dominate
in some cases over radiation and has been studied intensively \citep*{bisnovatyi_1977,shakura_1978,goldman_1995,agol_2001} with substantially
different conclusions (not least due to the underlying theoretical models): the contribution of convection to the overall energy 
transport regions ranges from $1/3$ \citep{shakura_1978} to being completely dominant \citep{bisnovatyi_1977} in radation 
pressure dominated disc regions. In gas pressure dominated regions, multiple solutions are found in the same range \citep{goldman_1995}. 
Also, recent 2-dimensional simulations by \citet{agol_2001} demonstrate that convective processes can release heat sufficiently 
fast to modify the vertical structure of the disc. It is therefore natural to consider the turbulence caused by convective 
motion as a possible candidate for viscosity. {First (semi-)\-analytical investigations were discouraging: they led to 
discs with masses comparable to or even exceeding the central black hole mass \citep{vila_1981,duschl_1989}, 
incompatible with the $\alpha$-viscosity description assumed in the models.} \citet{ruden_1988} and \citet{ryu_1992} studied convective
instabilities in thin gaseous discs and confirmed that angular momentum transport can be supported by convective turbulence.
\citet{goldman_1995} investigated accretion discs where viscosity is given by convection solely and where the energy transport is maintained by radiation and convection. They found the resulting viscosity being too low by a factor of $10$ to $100$, but could not draw final conclusions due to their limited disc model.

\textit{Magneto-rotational instability.} The magneto-rotational instability (MRI) was first noticed in a non-astrophysical context
by \citet{velikhov_1959} and \citet{chandrasekhar_1960}. More than 30 years later, \citet{balbus_1991,balbus_1998} established  that weak magnetic fields can substantially alter the
stability character of accretion discs, giving rise to a generic and efficient angular momentum transport. Today, the MRI is considered
as the primary candidate for the viscosity in astrophysical accretion flows. {Modern computational facilities allow
the study of angular momentum transport in magnetized discs in 3-dimensional MHD codes, which basically can be used to calibrate the
$\alpha$- or $\beta$-viscosity parameter} \citep[see, e.\,g.,][for a review]{balbus_2005}.
The key problem therein is the non-trivial dependency of $\alpha$ on various physical and numerical parameters of the simulations.
Recently, \citet*{pessah_2007} presented a scaling law which allows to disentangle physical and numerical influences. The general question if the MRI
effects can be translated into an $\alpha$- or $\beta$-type {viscosity
remains} to be answered \citep*[see, e.\,g.,][]{pessah_2008}. From the wealth of results obtained so far, it seems likely that the MRI alone cannot account for the viscosity in astrophysical discs
\citep{begelman_2007,king_2007}. For example, current results face a discrepancy of at least one order of magnitude between the viscosities
generated by the MRI and those inferred from observations \citep[see also][for a further discussion]{lesur_2007}.
A particular problem of the MRI are the so-called \emph{dead zones}, where the growth rate of magneto-rotational instabilities is
strongly suppressed and the turbulence induced by magnetic effects diminishes \citep{gammie_1996}. Although the implications of
MRI dead zones are discussed mostly for protoplanetary discs \citep*[see, e.\,g.,][for an overview]{reyes-ruiz_2003,brandenburg_2008}, the overall problem of a vanishing viscosity applies to accretion discs in general.

\medskip%
{Thus, it is not yet clear whether one of these candidates or a combination of them is responsible for generating the viscosity in astrophysical
discs.} One important step therefore is to {study} the effect of convective turbulence \emph{in combination} with other
contributors. \citet{goldman_1995} stressed the need for a convective disc model where the vertical structure is
calculated self-consistently in order to quantify better the convective turbulence and the energy transport in the disc.

In this paper, we construct a model of a black hole accretion disc where we
calculate the effect of convection in a self-consistent way by means of the mixing-length theory. Hereby, the total viscosity
is given by convection plus a supporting $\beta$-viscosity, accounting for turbulence due to differential rotation and allowing
for potential self-gravitating effects. Energy transport in the vertical direction is provided by radiation and convection simultaneously,
which allows to derive the strength of the convective viscosity within the model. The details of the model are given in Sect.~\ref{sec_model}; in Sect.~\ref{sec_results},
we present and analyse our results for various central black hole masses, accretion rates and values of the underlying $\beta$-viscosity.
Section~\ref{sec_discussion} is devoted to discussion and conclusion.
\section{Model setup}\label{sec_model}
We calculate accretion disc models where both the viscosity and the transport of energy is supported by convective processes,
in addition to an underlying $\beta$-viscosity and to radiative energy transport.
Hereby, convection is treated in the framework of the mixing-length theory. The disc
is assumed to be geometrically thin in order to allow for a $1+1$-dimensional treatment of the equations. 
We use a cylindrical coordinate system with planar radial coordinate $s$, vertical coordinate $z$ and
true radius $r=\sqrt{s^2+z^2}$. The disc geometry is determined by an inner and an outer radius, $s_\textup{i}$ and $s_\textup{o}$,
and the disc's thickness $h=h(s)$ from the mid-plane. 

The turbulent viscosity, caused by convective processes, is generally given by
\begin{equation} 
\nu_\textup{conv} = \xi v_\textup{conv} l_\textup{conv}\,
\end{equation}
with $v_\textup{conv}$ and $l_\textup{conv}$ being the turbulent velocity of the convective elements and the convective lengthscale over
which they diffuse, respectively. The factor $\xi$ is of the order of unity and depends on the degree of isotropy of convection in the considered direction.
For simplicity, we assume isotropy in this investigation (i.\,e., $\xi=1/3$). We identify the convective lengthscale $l_\textup{conv}$ with
the mixing-length $l_\textup{m}$, which will be defined later. We include other sources of viscosity (differential rotation, MRI, \ldots) by assuming a
permanently supporting viscosity to be present in the disc, parametrized by a standard $\beta$-ansatz:
\begin{equation} 
\nu_\beta = \beta s^2 \omega\,,\quad \beta \ll 1\,.\label{eqn_nu_beta}
\end{equation} 
{Here, $\omega$ stands for the angular velocity}. The total viscosity $\nu$ is then given by a combination of these two contributors,
\begin{equation} 
\nu = \nu_\textup{conv} + \nu_\beta\,.
\end{equation} 
\subsection{Radial structure}
For the calculation of the radial structure, we introduce
\begin{equation} 
\Psi = \int_{0}^{h} \nu \rho\,dz\,.\label{eqn_def_psi}
\end{equation}
{in analogy to the disc's surface density $\Sigma$,
\begin{equation} 
\Sigma = \int_{0}^{h} \rho\,dz\,.
\end{equation}}
Only in the special case of $\nu=\mbox{const.}$ can we rewrite~\eqref{eqn_def_psi} to
$\Psi = \nu \Sigma$. In all other cases, we apply the mean value theorem to define an
average value $\nu^{\star}$ such that
\[
\Psi = \nu^{\star} \Sigma
\]
The individual contributors $\nu^{\star}$ and $\Sigma$ remain unknown from the radial structure equations only,
but are determined by the vertical structure equations (Sect.~\ref{sec_vertical_structure_equations}).\pagebreak

The radial structure is determined by the conservation of mass, momentum, angular momentum and energy.
The corresponding equilibrium equations are
\begin{eqnarray} 
\dot{M}  &=& -4 \pi s v_s \Sigma\,,\label{eqn_continuity_equation}\\[3mm]
\omega^2 &=& \frac{g_s}{s}\,,\label{eqn_momentum}\\
2 \Psi &=&- \frac{\dot{M} \omega}{2\pi s (\partial \omega/\partial s)} \cdot \Bigl(1-\sqrt{s_\textup{i}/s}\Bigr)\,,\label{eqn_ang_momentum}\\
2 F  &=& - \frac{\dot{M} s \omega (\partial \omega/\partial s)}{2\pi} \cdot \Bigl(1-\sqrt{s_\textup{i}/s}\Bigr)\,.\label{eqn_energy}
\end{eqnarray}
{$\dot{M}$ denotes the (constant) accretion rate, $v_s$ the accretion velocity with $v_s<0$ for inflowing material, $g_s$ the graviational acceleration in radial direction, and
$F$ the heat flux, integrated in vertical direction.
We apply} the standard free-fall boundary condition \citep{shakura_1973,novikov_1973} in~\eqref{eqn_ang_momentum} and~\eqref{eqn_energy}, 
implying a vanishing torque at the disc's inner radius $s_\textup{i}$. The momentum equation~\eqref{eqn_momentum} is simplified by the
assumption of local equilibrium of the graviational attraction and the centrifugal repulsion in the radial direction. Therein, the gravitational acceleration
is assumed to be given by the monopole approximation \citep{mineshige_1997}, assuming a Pseudo-Newtonian gravitational potential~\citep{paczynski_1980}:
\begin{equation} 
g_s = \frac{G (M_\textup{c} + M_\textup{d}(s))}{(r-r_\textup{{S}})^2} \cdot \frac{s}{r}\,.\label{eqn_g_s_monopole}
\end{equation} 
{The Schwarzschild radius $r_\textup{S}$ is given by $2GM_\textup{c}/c^2$,
where $M_\textup{c}$ denotes the mass of the central black hole. The enclosed disc mass at radius $s$ is calculated via}
\begin{equation} 
M_\textup{d}(s) = \int_{s_\textup{i}}^{s} 4\pi s' \Sigma\,ds'\,.\label{eqn_Mdisc}
\end{equation} 
\subsection{Vertical stratification}\label{sec_vertical_structure_equations}
\subsubsection{Structure equations}
In analogy to \citet{cannizzo_1988}, \citet{hofmann_2005} and \citet{vehoff_2005}, we adopt the energy
flux {at height $z$ as the independent coordinate for the vertical integration:
\begin{equation} 
F_z = \int_{0}^{z} \dif{F_z}{z}\,dz\,, \qquad F_z(z=h) = F\,.
\label{eqn_def_heat_flux_at_z}
\end{equation} }
Additionally, we introduce the surface density at height $z$,
\begin{equation} 
\Sigma_z = \int_{0}^{z} \rho\,dz\,, \qquad \Sigma_z(z=h) = \Sigma\,,
\label{eqn_def_surface_density_at_z}
\end{equation} 
and
\begin{equation} 
\psi = \int_{0}^{z} \nu \rho\,dz\,, \qquad \psi(z=h) = \Psi\,.
\label{eqn_def_psi_at_z}
\end{equation} 
Neither $\Sigma$ nor $h$ are known \emph{a priori} -- they will be a result of the vertical integration.
{The equations for the vertical structure of the disc are given as follows:}

\begin{eqnarray} 
\dif{z}{F_z}&=&\frac{1}{\rho \nu s^2 \left(\dif{\omega}{s}\right)^2}\,,\label{eqn_vert_strat_z_0}\\
\dif{T}{F_z}&=&-\frac{1}{\rho \nu s^2 \left(\dif{\omega}{s}\right)^2} \cdot \nonumber\\
&&\cdot \left\{(1-\zeta) \frac{3\kappa \rho F_z}{4acT^3} + \zeta \frac{g_z (4-3\gamma) }{\gamma c_\textup{p}} \right\}\,,\label{eqn_vert_strat_T_0}\\
\dif{\Sigma_z}{F_z}&=&\frac{1}{\nu s^2 \left(\dif{\omega}{s}\right)^2}\,,\label{eqn_vert_strat_Sigma_0}\\
\dif{p}{F_z} &=& - \frac{g_z}{\nu s^2 \left(\dif{\omega}{s}\right)^2}\,.\label{eqn_vert_strat_p_0}
\end{eqnarray} 
Here, $c_\textup{p}$ stands for the isobaric specific heat capacity and $\gamma=p_\textup{gas}/p$.
{\eqref{eqn_vert_strat_z_0} is a simple inversion of the local energy production by viscous dissipation,
$\partial F_z/\partial z = \rho \nu s^2 (\partial\omega/\partial s)^2$. The temperature stratification \eqref{eqn_vert_strat_T_0} results from accounting for the energy transport by
radiation and convection \citep{cox_1968} and using~\eqref{eqn_vert_strat_z_0}}. Relating the two terms in \eqref{eqn_vert_strat_T_0} with the radiative and adiabatic
gradients $\nabla_\textup{rad}$ and $\nabla_\textup{ad}$, it becomes clear that the variable $\zeta$ describes
the relative contribution of the convective energy transport to the total energy transport. Its value depends on
the \emph{local} physical conditions at position $(s,z)$ in the disc and can be calculated numerically, see Sect.~\ref{sec_mlt}.
{Combining~\eqref{eqn_vert_strat_z_0} with the definition of the surface density at height $z$~\eqref{eqn_def_surface_density_at_z} leads to~\eqref{eqn_vert_strat_Sigma_0}.
Assuming hydrostatic equilibrium and again using~\eqref{eqn_vert_strat_z_0} gives the last differential equation~\eqref{eqn_vert_strat_p_0}
for the pressure stratification.}

Using~\eqref{eqn_vert_strat_T_0},~\eqref{eqn_vert_strat_p_0}, and the equation of state,
\begin{equation} 
p = p_\textup{gas} + p_\textup{rad} = \frac{\rho k_\textup{B} T}{\mu m_\textup{H}} + \frac{4 \sigma_\textup{SB}}{3c} T^4\,,
\label{eqn_of_state}
\end{equation} 
we transform~\eqref{eqn_vert_strat_p_0} into an equation for the mass density $\rho$:
\begin{eqnarray} 
\dif{\rho}{F_z} &=& \frac{\mu m_\textup{H}}{k_\textup{B} \nu s^2 \left(\dif{\omega}{s}\right)^2} \cdot
      \left[-\frac{g_z}{T} + \left(\frac{p}{\rho T^2} + \frac{4\sigma_\textup{SB} T^2}{\rho c} \right) \cdot\right.\nonumber\\
&&\qquad\left. \cdot \left\{(1-\zeta) \frac{3\kappa \rho F_z}{4acT^3}
                      + \zeta \frac{g_z (4-3\gamma)}{\gamma c_\textup{p}} \right\} \right]\,.\label{eqn_vert_strat_rho_0}
\end{eqnarray} 
For the numerical solution of the vertical stratification, the opacity $\kappa = \kappa(\rho, T)$ is calculated from a combination of tabulated values and
interpolation formulae, see Sect.~\ref{sec_def_kappa}. In analogy to $g_s$, the gravitational acceleration in vertical direction $g_z$ is provided
by the mono\-pole approximation:
\begin{equation} 
g_z = \frac{G (M_\textup{c} + M_\textup{d}(s))}{(r-r_\textup{{S}})^2} \cdot \frac{z}{r} + 4 \pi G \Sigma_z\,.\label{eqn_g_z_monopole}
\end{equation}
In \eqref{eqn_g_z_monopole}, the second term stands for the local gravitational attraction, which becomes important in the self-gravitating regime.
\subsubsection{Adaptation of the mixing length theory}\label{sec_mlt}
In order to solve the vertical structure equations, we apply the mixing-length theory \citep{boehm_1958} as formulated in \citet{cox_1968}.
The mixing-length theory expresses the efficiency of the convective energy transport relative to the radiative transport processes
by the variable $\zeta$, where $0\leq \zeta\leq 1$. A vanishing $\zeta$ implies no convective transport, while in the case $\zeta=1$ all energy is
transported by convection. {Following \citet{cox_1968}}, its value can be calculated from the cubic equation
\begin{equation} 
\zeta^{1/3} + B \cdot \zeta^{2/3} + a_0 B^2 \zeta - a_0 B^2 =0\,,
\label{eqn_zeta_disc}
\end{equation} 
with a numerical factor $a_0 = 9/4$ and further quantities defined as
\begin{eqnarray*}
B   & = & \left[\frac{A^2}{a_0} \cdot (\nabla_\textup{rad} - \nabla_\textup{ad})\right]^{1/3}\,,\\
A^2 & = & \frac{Q \cdot (c_\textup{p} \kappa g_z)^2 \rho^5 l_\textup{m}^4}{288 a^2 c^2 p T^6}\,,\\
Q & = & \frac{4-3\gamma}{\gamma}\,,\\
c_\textup{p} & = & \frac{\mathfrak{R}}{\mu} \cdot \frac{32 - 24 \gamma - 3 \gamma^2}{2 \gamma^2}\,,\\
\nabla_\textup{rad} & = & \frac{3 \kappa \rho \lambda_\textup{p} F_z}{4ac T^4}\,,\\
\nabla_\textup{ad}  & = & \frac{8 - 6 \gamma}{32 - 24 \gamma - 3\gamma^2}\,,\\
l_\textup{m} & = & \min(\lambda_\textup{p},h)\,,\\
\lambda_\textup{p} & = & \frac{p}{g_z \rho}\,.
\end{eqnarray*}
The mixing-length $l_\textup{m}$ is usually of the order of the pressure scale height $\lambda_\textup{p}$.
However, in analogy to the stellar case, it is limited by simple geometric effects.
While in the stellar case, it usually cannot exceed the actual radial distance from the centre due to symmetry requirements,
we adopt the actual height $h=h(s)$ of the disc as upper limit. {In doing so, convective elements are allowed to travel across the disc mid plane,
which overrides the symmetry of the disc. Also, it removes the strict upper barrier (i.\,e., the disc surface) for the convective elements.}
Nevertheless, it provides a simple method of taking into account {overshooting} effects and a more realistic, smooth transition between the disc and the atmosphere. We note that the results
differ only slightly {for a more restrictive definition $l_\textup{m}=\min(\lambda_\textup{p},h-z,z)$}, so that our conclusions {do not depend} on this assumption.

From the above definitions, the cubic equation~\eqref{eqn_zeta_disc} is solved numerically. Subsequently, the convective viscosity is calculated
from {\citep{cox_1968}}
\begin{equation} 
v_\textup{conv} = c_\textup{s} \cdot \frac{Q^{1/2} \alpha_{l_\textup{m}}}{2\sqrt{2} \Gamma_1^{1/2}}
    \left(\frac{\nabla_\textup{rad} - \nabla_\textup{ad}}{a_0 A}\right)^{1/3} \zeta^{1/3}\,,\label{eqn_vconv_disc}
\end{equation} 
where $c_\textup{s}$ denotes the sound speed at the actual coordinate $(s,z)$ in the disc. In the non-relativistic
regime, it is given by
\begin{equation} 
c_\textup{s} = \sqrt{\Gamma_1\,p/\rho}\,.\label{eqn_cs_disc}
\end{equation} 
The constant $\Gamma_1$ stands for the polytropic index, which is given
by $5/3$ in the case of a non-relativistic, ideal gas. The parameter $\alpha_{l_\textup{m}}$ relates the typical
distance traveled by the convective elements to the pressure scale
height:
\begin{equation} 
\alpha_{l_\textup{m}} = \frac{l_\textup{m}}{\lambda_\textup{p}} \leq 1\,.\label{eqn_alpha_disc}
\end{equation}%
The derivation of the mixing-length theory assumes a purely subsonic motion of the convective elements,
$v_\textup{conv}\leq c_\textup{s}$. This inequality cannot be assured by the definition of the
convective velocity in~\eqref{eqn_vconv_disc}. {Hence, in the case~\eqref{eqn_vconv_disc} leads to values $v_\textup{conv} > c_\textup{s}$,
we follow \citet{cox_1968} and set manually $v_\textup{conv}=c_\textup{s}$. This, in turn, invalidates equation~\eqref{eqn_zeta_disc} for $\zeta$.
Instead, $\zeta$ must be calculated from~\eqref{eqn_vconv_disc} with $v_\textup{conv}=c_\textup{s}$:}
\begin{equation} 
\zeta = \frac{8 \sqrt{8} \Gamma_1^{3/2} a_0^{3/2} A}{Q^{3/2} \alpha_{l_\textup{m}}^3 (\nabla_\textup{rad}-\nabla_\textup{ad})}\,.\label{eqn_zeta_disc_sonic_limit}
\end{equation} 
{This correction leads to lower values of $\zeta$ and therefore to an effective decrease of the convective energy transport and of the convective
viscosity $\nu_\textup{conv}$.}
\subsubsection{Boundary conditions}\label{sec_boundary_conditions}
We define the boundary conditions for the four equations \eqref{eqn_vert_strat_z_0}--\eqref {eqn_vert_strat_Sigma_0}, \eqref{eqn_vert_strat_rho_0}
at either the disc's mid plane (``mp'', $F_z=0$) or surface (``eff'', $F_z=F$):
\begin{eqnarray} 
z_\textup{mp}      &=& z(F_z=0)\ =\ 0\,,\label{eqn_vert_strat_bc_z}\\[2mm]
T_\textup{eff}     &=& T(F_z=F)\ =\ \left(\frac{F}{\sigma_\textup{SB}}\right)^{1/4}\,,\label{eqn_vert_strat_bc_T}\\[3mm]
\Sigma_{z,\textup{mp}} &=& \Sigma_z(F_z=0)\ =\ 0\,,\label{eqn_vert_strat_bc_Sigma}\\[4mm]
\rho_\textup{eff}  &=& \rho(F_z=F) \ =\ \rho_\textup{eff,input}\,.\label{eqn_vert_strat_bc_rho}
\end{eqnarray} 
The boundary condition on $\rho$ cannot be determined from the radial structure equations or from simple geometric arguments.
It rather requires us to define an atmosphere above the disc, which allows us to determine the density $\rho_\textup{eff,input}$ consistently
with the height $h$, the surface density $\Sigma$ and the effective temperature $T_\textup{eff}$. Details about the atmosphere will be given in
Sect.~\ref{sec_atmosphere_model}.
\subsubsection{Opacity $\kappa$}\label{sec_def_kappa}
The opacity has a rather strong influence on the computation of the vertical structure. Therefore, we refrain from simple models such as
pure electron scattering or Kramer's law. We rather use a combination of tabulated values and interpolation
formulae to calculate the opacity for a broad range of temperatures and densities. The presence of a disc and an atmosphere implies
the calculation of Rosseland and Planck opacities.
\subparagraph*{Tabulated values}
Given that we want to cover a large domain in temperature and density, multiple sources are included in our model. In the high-temperature
limit, we adopt the tables from the TOPS project \citep{tops_www}. We compile tables for Rosseland and Planck opacities in the range of
\begin{eqnarray*}
\log \rho&=&\left[ -12.5 \ldots +10.5 \right]\,,\\
\log T&=&\left[ +4.5 \ldots +9.1 \right]\,.
\end{eqnarray*}
All values are given in cgs-units. The number of data points is $47$ on an equidistant scale for $\log \rho$ and $41$ for $\log T$, respectively.

In the low-temperature regime, we include the Ferguson opacities \citep{ferguson_2005,ferguson_www}. We compile Rosseland and Planck
opacity tables in the range of
\begin{eqnarray*}
\log R&=&\left[ -8.0 \ldots +1.0 \right]\,,\\
\log T&=&\left[ +2.7 \ldots +4.5 \right]\,,
\end{eqnarray*}
with $R=\rho/T_6^3$ ($T_6 = T/10^6$). These ranges correspond to minimum and
maximum mass densities of $\log \rho=-17.9$ and $+6.5$, with a resolution of $19$ equidistant points in $\log R$ and $85$ in $\log T$.

We choose identical chemical abundances for the TOPS and Ferguson opacities with mass fractions $X=0.7$, $Y=0.28$, $Z=0.02$, and
the chemical mixture of~\citet{grevesse_1998}.
\subparagraph*{Analytic interpolation formula}
Opacities outside the ranges given above are calculated using
an analytic interpolation formula \citep[Gail, priv.\,comm.; for a similar approach, see also][]{bell_1994}
for Rosseland opacities. Thus, strictly speaking, this interpolation formula is valid only in the optically thick regions.

\begin{eqnarray} 
\frac{1}{\kappa_\textup{in}}&=&
\left[\frac{1}{\kappa_\textup{ice}^4} +
\frac{T_0^{10}}{T_0^{10} + T^{10}}\cdot
\frac{1}{\kappa_\textup{ice, evap}^4 + \kappa_\textup{dust}^4}\right]^{1/4}\nonumber\\
&& + \left[\frac{1}{\kappa_\textup{dust, evap}^4 +
\kappa_\textup{mol}^4 + \kappa_\textup{H$^{-}$}^4}
+\frac{1}{\kappa_\textup{atom}^4 + \kappa_\textup{e$^{-}$}^4}\right]^{1/4}
\label{eqn_def_kappa_interpol}
\end{eqnarray} 
The individual contributors $\kappa_l$ are approximated by
\begin{equation} 
\kappa_l = \kappa_{0,l} \cdot T^{\kappa_{T,l}} \cdot \rho^{\kappa_{\rho,l}}
\end{equation} 
and are compiled in Table~\ref{tab_kappa_interpol}. {The temperature $T_0$ parametrizes the transition
between atomic/molecular ice and gas and is set to $T_0=3000\,\textup{K}$.}
Note that the definitions of {$T_0$ and of} the individual contributors are such that the {full interpolation
formula~\eqref{eqn_def_kappa_interpol}} fits {the values obtained from experiments and numerical calculations};
they cannot be used on their own as a physical descriptions of the corresponding processes.
\begin{table}
\caption{Interpolation of the opacity: set of parameters (in cgs-units).}\label{tab_kappa_interpol}
{\tabcolsep3pt\begin{tabular}{lcccc}
\hline
\rule[1mm]{0mm}{2mm}\textbf{Contributor $\mathbf{l}$} & \textbf{Symbol} &
$\mathbf{\kappa_{\textup{l},0}}$
    & $\mathbf{\kappa_{\textup{l},\rho}}$ & $\mathbf{\kappa_{\textup{l},T}}$ \\ \hline
\rule[1mm]{0mm}{2mm}Dust with ice mantles & $\kappa_\textup{ice}$ & $2.0 \cdot 10^{-4}$ & $0$ & $2$\\
Evaporation of ice & $\kappa_\textup{ice, evap}$ & $1.0 \cdot 10^{16}$ & $0$ & $-7$\\
Dust particles & $\kappa_\textup{dust}$ & $1.0 \cdot 10^{-1}$ & $0$ & $1/2$\\
Evaporation of dust particles & $\kappa_\textup{dust, evap}$ & $2.0 \cdot 10^{81}$ & $1$ & $-24$\\
Molecules & $\kappa_\textup{mol}$ & $1.0 \cdot 10^{-8}$ & $2/3$ & $3$\\
Negative hydrogen ion & $\kappa_\textup{H$^{-}$}$ & $1.0 \cdot 10^{-36}$ & $1/3$ & $10$\\
Bound-free, free-free-transitions & $\kappa_\textup{atom}$ & $1.5 \cdot 10^{20}$ & $1$ & $-5/2$\\
Electron scattering & $\kappa_\textup{e$^{-}$}$ & $0.348$ & $0$ & $0$\\ \hline
\end{tabular}}
\end{table}
\subparagraph*{Opacity mixture}
For smooth transitions and a broad coverage in the $T$-$\rho$ range, we use a combination of the tabulated opacities
(TOPS, Ferguson) and the interpolation formula~\eqref{eqn_def_kappa_interpol}. The transition between the TOPS and the Ferguson opacities
takes place at $\log T=4.5$, modeled by a linear interpolation of the opacities from both sources in the range $\log T = [4.0 \ldots 5.0]$.

\begin{figure*}
\includegraphflex[width=0.48\textwidth,clip]{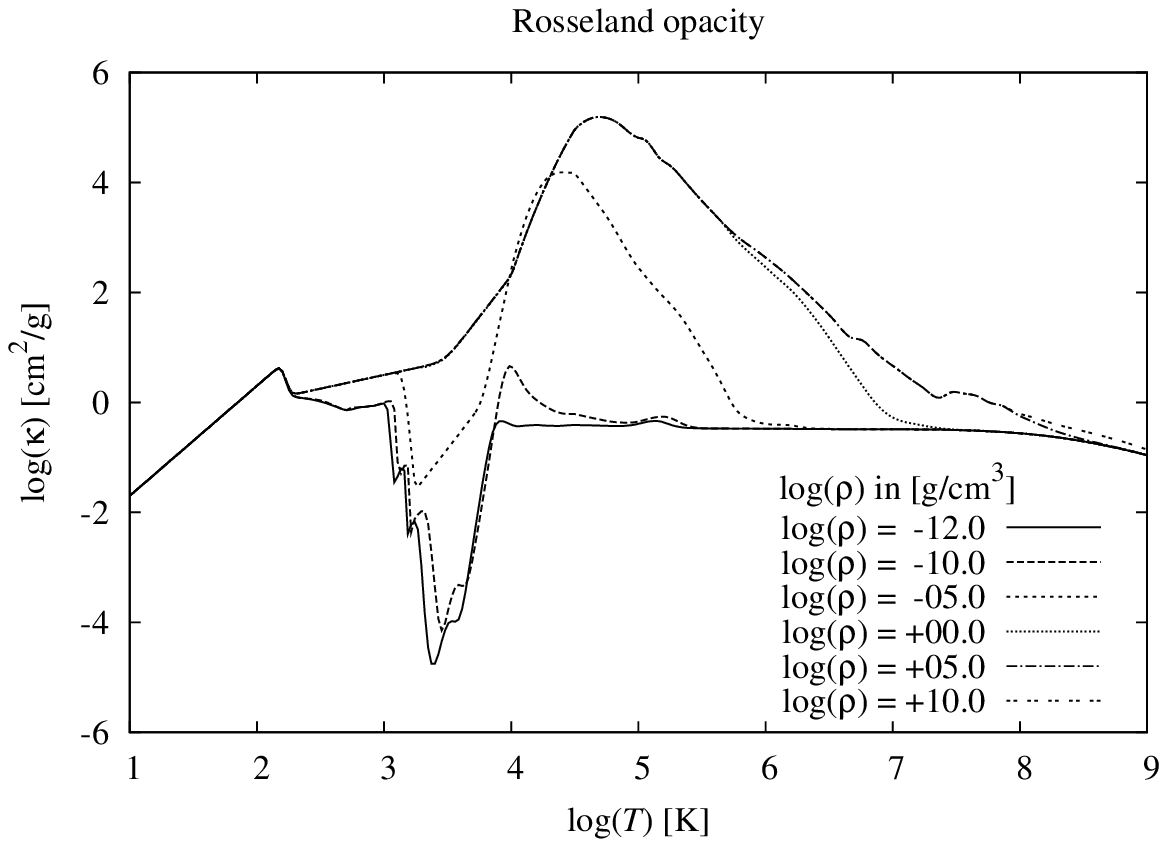}\hfill%
\includegraphflex[width=0.48\textwidth,clip]{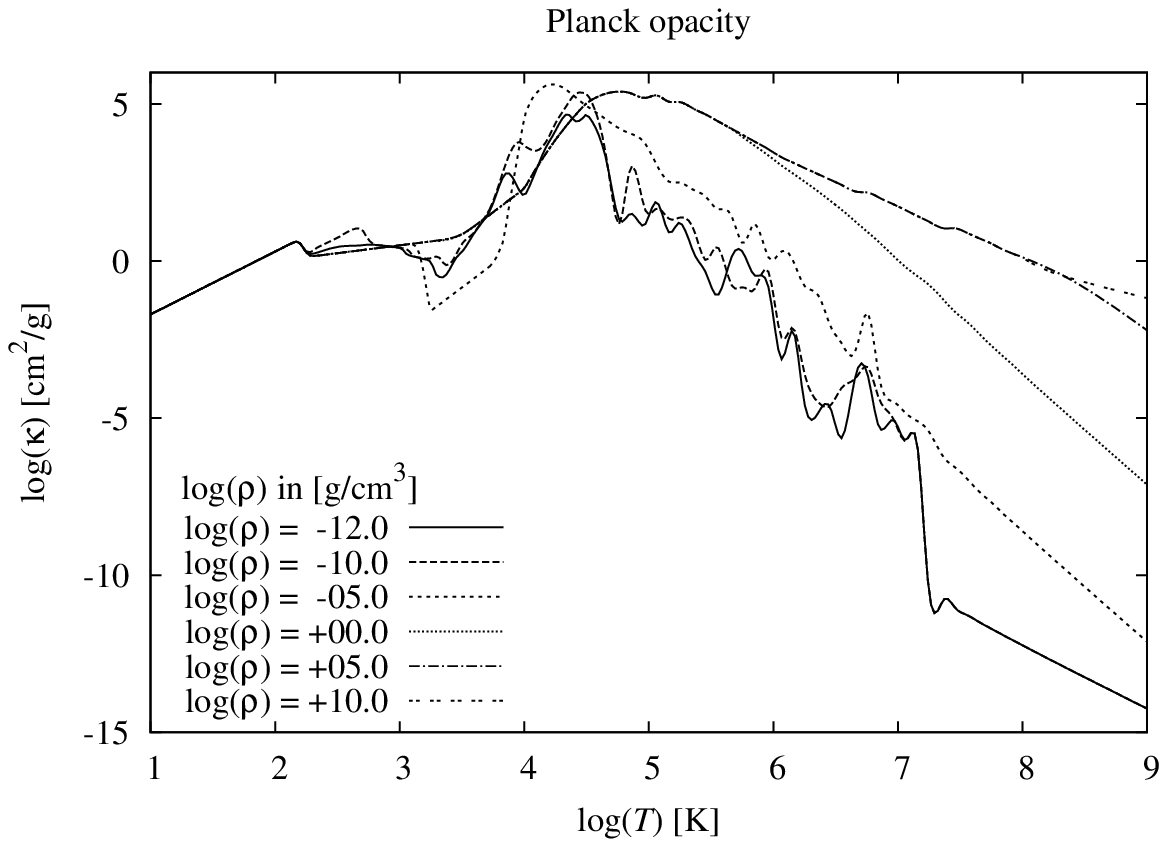}\\
{\small\textbf{(a)}}\rule{0.48\textwidth}{0mm}{\small\textbf{(b)}}\hfill{\small \phantom{x}}%
\caption{Temperature dependency of the \textbf{(a)} Rosseland and \textbf{(b)} Planck opacities for certain densities}\label{fig_kappa_R_P}
\end{figure*}

At the ``outer'' boundaries of the TOPS- and Ferguson-opacities, we use the same kind of linear transition in a range of
$\Delta\log\rho = 1$ and $\Delta\log T = 1$ between the tabulated values and the interpolation formula~\eqref{eqn_def_kappa_interpol}.
The resulting opacities are defined on a $\log T$--$\log\rho$ grid with 150 data points in each direction and
\begin{eqnarray*}
\log \rho&=&\left[ -15.0 \ldots +10.0 \right]\,,\\
\log T&=&\left[ +1.0 \ldots +9.0 \right]\,,
\end{eqnarray*}
which is sufficient for our purposes. {Figures~\ref{fig_kappa_R_P}a,b display the resulting opacities as a function of temperature for certain densities.}
\subsubsection{Atmosphere}\label{sec_atmosphere_model}
The only purpose of the atmosphere is to provide a value for the mass density at the surface of the accretion disc at each radial position $s$,
which is consistent with the actual effective temperature, geometrical height and surface density. Therefore, it is sufficient to calculate a
simple grey atmosphere in the Milne-Eddington way, where the temperature distribution is given as a function of the optical depth $\tau$ by
\begin{equation} 
T(\tau)^4 = C_{1,\textup{atm}} T_\textup{eff}^4 \cdot ( \tau + C_{2,\textup{atm}} )\,.\label{eqn_atm_temperature_layering}
\end{equation} 
The constants $C_{1,\textup{atm}}$ and $C_{2,\textup{atm}}$ depend on the transition point $\tau_\textup{eff}$
between the atmosphere (optically thin) and the disc (optically thick) and the final value for the temperature
at the ``upper'' end of the atmosphere ($\tau \ll 1$). We use the common value of
$ T^4(\tau=0) = (1/2) \cdot T_\textup{eff}^4$, but allow the transition to take place at $\tau_\textup{eff} = 1$ (instead
of the common value $2/3$) for a simple reason: in the optical thin atmosphere, the equation of state~\eqref{eqn_of_state}
is modified such that the radiation pressure term tends to zero for $\tau \to 0$. With the approximate expression for the
radiation pressure in an optical thin medium{~\citep[see, e.\,g.,][for a detailed discussion]{artemova_1996}, the atmospheric equation of state becomes
\begin{equation} 
p = p_\textup{gas} + p_\textup{rad,atm} = \frac{\rho k_\textup{B} T}{\mu m_\textup{H}} + \frac{4 \sigma_\textup{SB}}{3c} \tau T^4\,.
\end{equation} 
A smooth transition of the pressure between the disc and the atmosphere requires $\tau_\textup{eff} = 1$, which implies
$C_{1,\textup{atm}} = 1/2$ and $C_{2,\textup{atm}} = 1$}. The remaining equations are given as follows: from the definition of the optical depth,
\[
d\tau = - \kappa \rho\,dz\,,
\]
we get an expression for $\partial z/\partial \tau$. The differential expression for the surface density
\[
d\Sigma_z = \rho\,dz
\]
is transformed into $\partial \Sigma_z/\partial \tau$. We choose the gas pressure as the fourth dependent variable and assume hydrostatic equilibrium,
$\partial p/\partial z = -\rho g_z$, to obtain the following set of differential equations for the structure of the atmosphere:
\begin{eqnarray} 
\dif{z}{\tau} &=& -\frac{1}{\rho \kappa}\ =\ -\frac{k_\textup{B} T}{\mu m_\textup{H} \kappa p_\textup{gas}}\,,\label{eqn_vert_strat_atm_z}\\
\dif{\Sigma_z}{\tau} &=& -\frac{1}{\kappa}\,,\label{eqn_vert_strat_atm_Sigma}\\
\dif{T}{\tau} &=& T_\textup{eff} \cdot \left(\frac{1}{2}\right)^{9/4} \cdot \left(\tau + 1\right)^{-3/4}\,,\label{eqn_vert_strat_atm_T}\\
\dif{p_\textup{gas}}{\tau} &=& \left( \frac{g_z}{\kappa} - \frac{4 \sigma_\textup{SB}}{3c} T_\textup{eff}^4
                                \cdot \left(\tau + \frac{1}{2} \right) \right)\,.\label{eqn_vert_strat_atm_pgas}
\end{eqnarray} 
{Equation~\eqref{eqn_vert_strat_atm_T} is derived from the temperature profile~\eqref{eqn_atm_temperature_layering}, and
\eqref{eqn_vert_strat_atm_pgas} is calculated from
\[
\dif{p_\textup{gas}}{\tau} = \dif{p}{\tau} - \dif{p_\textup{rad,atm}}{\tau} = - \rho g_\textup{z} \dif{z}{\tau} - \frac{4 \sigma_\textup{SB}}{3c} \dif{}{\tau} \left(\tau T^4\right)\,.
\]}
The corresponding boundary conditions need to be set at either the lower boundary (i.\,e., at the disc surface, corresponding to $\tau = \tau_\textup{eff} = 1$) or the upper
boundary (``up'', corresponding to $\tau = \tau_\textup{up} \ll 1$). Three of these boundary conditions are provided by the solution of the vertical disc structure:
\begin{eqnarray} 
h &=& z(F_z=F)\ =\ z(\tau=\tau_\textup{eff})\,,\label{eqn_vert_strat_atm_bc_z}\\
\Sigma &=& \Sigma_z(F_z=F)\ =\ \Sigma_z(\tau=\tau_\textup{eff})\,,\label{eqn_vert_strat_atm_bc_Sigma}\\
T_\textup{eff} &=& T(F_z=F)\ =\ T(\tau=\tau_\textup{eff})\,.\label{eqn_vert_strat_atm_bc_T}
\end{eqnarray} 
The fourth boundary condition on the gas pressure has to be set at the upper boundary of the atmosphere, since
we want to calculate a consitent value of $p_\textup{gas}$ (i.\,e., $\rho$) at the disc surface. We define a constant minimum
value for the mass density
\begin{equation} 
\rho_\textup{up} = \rho(\tau=\tau_\textup{up}) = \mbox{const.}
\label{eqn_vert_strat_atm_bc_pgas}
\end{equation} 
and calculate the corresponding value $p_\textup{gas,up}$ at every radial position from $T_\textup{up}=T(\tau_\textup{up})$ and $\rho_\textup{up}$ (see Table~\ref{tab_parameters_settings} {for
the numerical values of $\rho_\textup{up}$ and $\tau_\textup{up}$}).
\subsection{Numerical solution}\label{sec_numerical_solution}
Thanks to the $1+1$-dimensional model, the radial equations decouple from the vertical structure and can be solved separately. The application of the
monopole approximation for the disc's self-gravity requires the enclosed disc mass $M_\textup{d}$ at radius $s$ to be known for solving the radial structure equations
(c.\,f. \eqref{eqn_continuity_equation}--\eqref{eqn_energy}, \eqref{eqn_g_s_monopole}). A priori, this is only the case at the inner disc radius, where $M_\textup{d}=0$.
{Due to the inner boundary condition, $\Psi$ and $F$ tend to zero for $s\to s_\textup{i}$ (c.\,f., \eqref{eqn_ang_momentum},~\eqref{eqn_energy}), which causes numerical problems when trying to solve the vertical stratification. Hence, we start the calculation close to the inner boundary, where $M_\textup{d}\ll M_\textup{c}$. First, we solve the radial structure equations.}
With the resulting values of $\Sigma$, $\Psi$ and $T_\textup{eff}$, the vertical structure
can be calculated numerically {in the second step}, allowing to update the enclosed disc mass~\eqref{eqn_Mdisc} and to proceed outwards in radial direction.

We use two separate methods to calculate the vertical stratification in the disc and the atmosphere. The disc equations are obviously more complicated to solve and as such they are more prone to numerical issues
like, e.\,g., steep gradients. We therefore apply a Henyey algorithm \citep*{henyey_1964} for solving the set of differential equations
in the disc. The Henyey method looks back on a successful history of applications in stellar structure and evolution codes, being
able to deal with steep gradients by its relaxation method nature. The atmospheric equations, however, are much easier to solve
and do not require a powerful, yet expensive, algorithm like the Henyey method. We apply a standard shooting algorithm to solve
the atmospheric structure in a simple and quick way. Details about the numerical methods are presented in \citet{heinzeller_2008}.

To determine the consistency of the numerical solution for the vertical stratification at each radius $s$, we iterate between the Henyey solver for the disc
and the shooting method for the atmosphere. Given an inital guess for $\rho_\textup{eff}$ and the boundary conditions \eqref{eqn_vert_strat_bc_z}--\eqref {eqn_vert_strat_bc_Sigma},
the former one provides values for $\Sigma$, $h$ and $T_\textup{eff}$, once the Henyey solver converged to the correct solution.
The latter one updates the input value $\rho_\textup{eff,disc}$ from the disc solution by solving the atmospheric stratification for the given $\Sigma$, $h$ and $T_\textup{eff}$
and the boundary conditions to $\rho_\textup{eff,atm}$. The combined solution is accepted for
\begin{equation} 
\left| \rho_\textup{eff,disc} - \rho_\textup{eff,atm} \right| \stackrel{!}{\leq}
    \epsilon \cdot \min\left\{\rho_\textup{eff,disc},\rho_\textup{eff,atm}\right\}\,,
\end{equation}
with the required accuracy $\varepsilon$ being defined in Table~\ref{tab_parameters_settings}.
\section{Results}\label{sec_results}
\begin{table}
\caption{Parameters and settings in the numerical model}\label{tab_parameters_settings}
{\tabcolsep3pt\begin{tabular}{lll} \hline
\rule[2mm]{0mm}{2.5mm}Central black hole mass & $M_\textup{c}$ & $1 M_\odot\ldots 100 M_\odot$\\
Accretion rate & $\dot{M}$ & $0.01 \dot{M}_\textup{E} \ldots 0.15 \dot{M}_\textup{E}$\\
Standard $\beta$-viscosity parameter & $\beta$ & $1\cdot10^{-3} \ldots 5\cdot 10^{-6}$\\
Corresponding $\alpha$ parameter & $\alpha$ & $3\cdot10^{-2} \ldots 2\cdot 10^{-4}$\\
Inner disc radius & $s_\textup{i}$ & $3 r_\textup{S}$\\
Outer disc radius & $s_\textup{o}$ & $500 r_\textup{S}$\\
Optical depth at upper end of atm. & $\tau_\textup{up}$ & $10^{-4}$\\
Density at upper end of atmosphere & $\rho_\textup{up}$ & $10^{-12}$\\[0.5ex]
Grid points in $s$ direction & $N_s$ & $100$\\
Default grid points in $z$-direction & $N_{z,\textup{ini}}$ & $100$\\
Maximum grid points in $z$-direction & $N_{z,\textup{max}}$ & $210$\\
Max. deviation of disc and atm. sol. & $\epsilon$ & $0.01$\\[0.5ex] \hline
\end{tabular}}
\end{table}
The results presented below were obtained for the set of parameters given in Table~\ref{tab_parameters_settings}.
Due to the free-fall inner boundary condition imposed on~\eqref{eqn_ang_momentum} and~\eqref{eqn_energy},
we start the radial calculation at {$s=2s_\textup{i}=6r_\textup{S}$ (see also Sect.~\ref{sec_numerical_solution})}.
From the values for the surface density obtained at $s=2s_\textup{i}$, we can estimate the enclosed disc mass
for $s_\textup{i}\leq s<2s_\textup{i}$, finding that its contribution is more than ten orders of magnitude smaller than the central mass
in all cases.

We further verified that in all cases the atmosphere contains almost no mass, compared to the vertical column of the underlying disc.
The atmosphere is thin ($z_\textup{min}\lessapprox h$) in in the innermost disc region, but expands up to $10 h$ in the outer regions due to a significantly
smaller gravitational attraction towards the disc mid plane.
\subsection{Disc properties of the standard disc model}\label{sec_standard_disc}
\begin{figure*}%
\begin{minipage}{0.47\textwidth}%
\includegraphflex[clip,width=\textwidth]{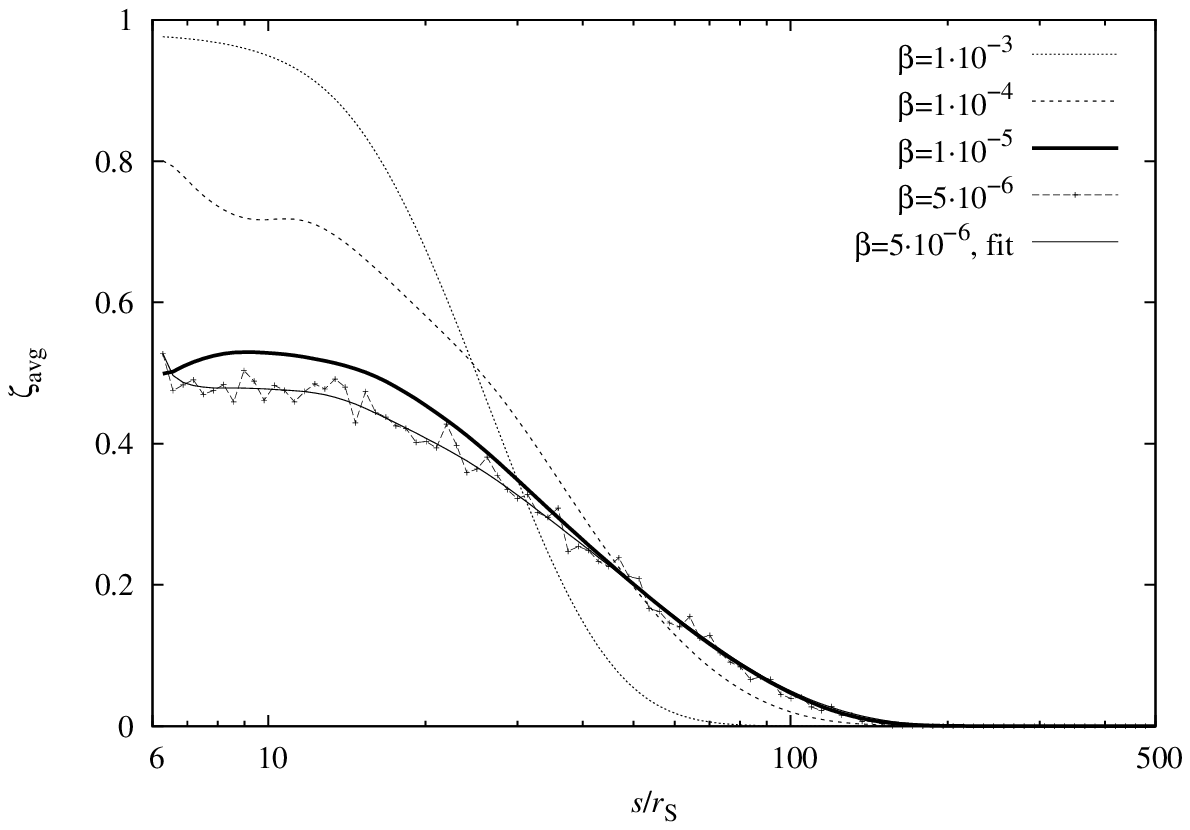}%
\end{minipage}\hfill%
\begin{minipage}{0.47\textwidth}%
\includegraphflex[clip,width=\textwidth]{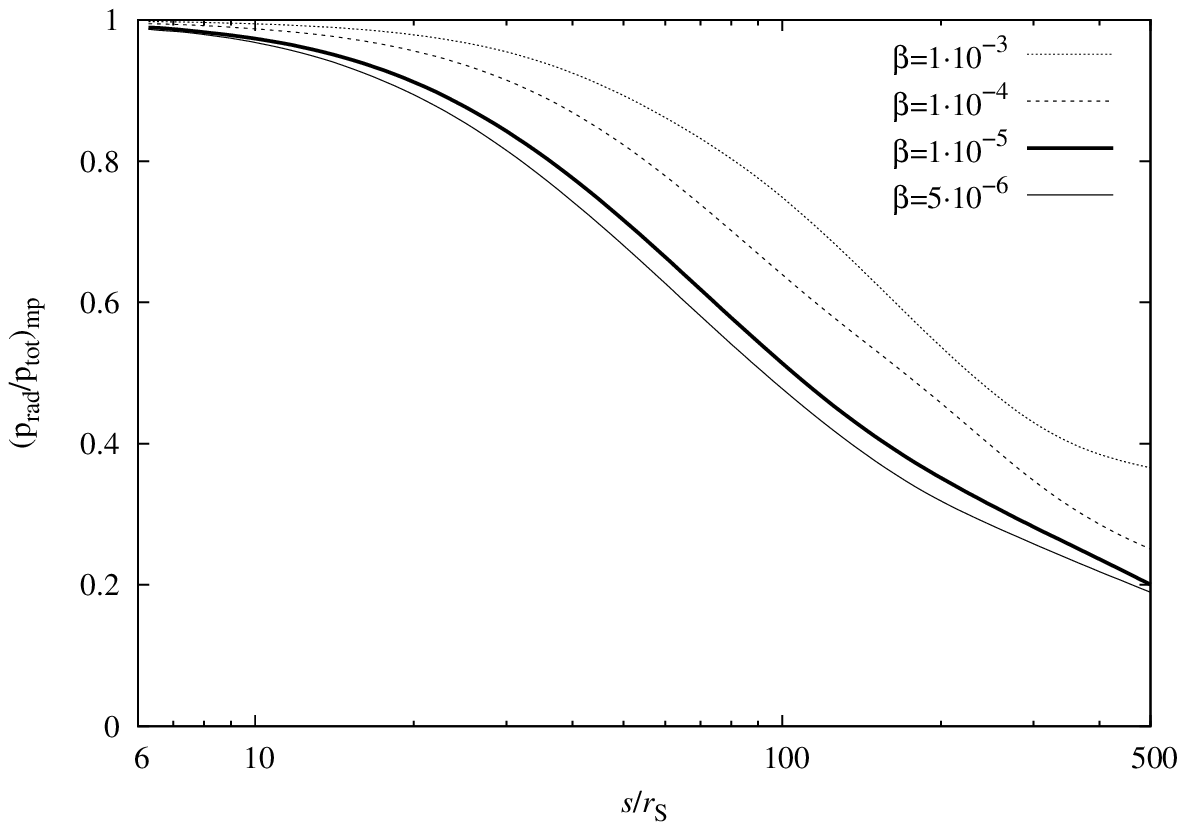}%
\end{minipage}\\%
\vskip-4ex{\small\textbf{(a)}}\rule{0.48\textwidth}{0mm}{\small\textbf{(b)}}\hfill{\small \phantom{x}}\\[1ex]%
\begin{minipage}{0.47\textwidth}%
\includegraphflex[clip,width=\textwidth]{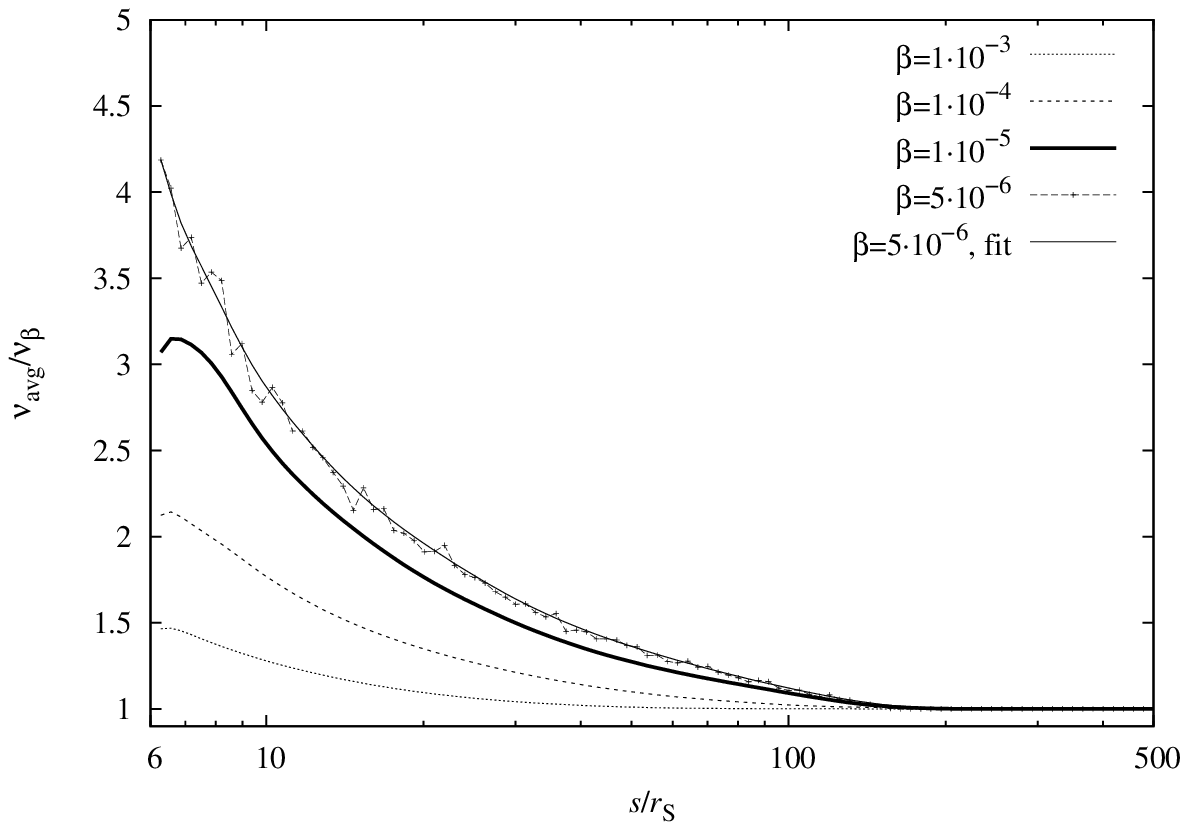}%
\end{minipage}\hfill%
\begin{minipage}{0.47\textwidth}%
\includegraphflex[clip,width=\textwidth]{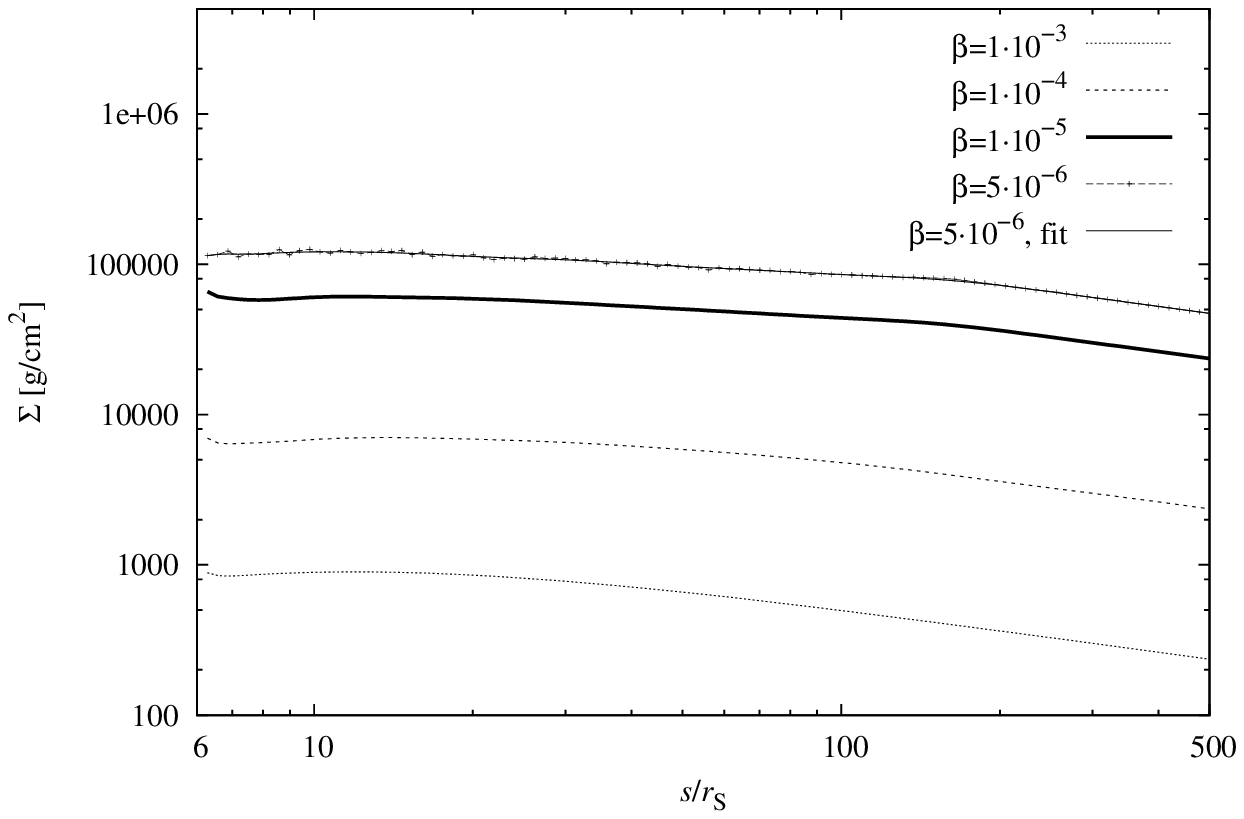}%
\end{minipage}\\%
\vskip-4ex{\small\textbf{(c)}}\rule{0.48\textwidth}{0mm}{\small\textbf{(d)}}\hfill{\small \phantom{x}}\\[1ex]%
\begin{minipage}{0.47\textwidth}%
\includegraphflex[clip,width=\textwidth]{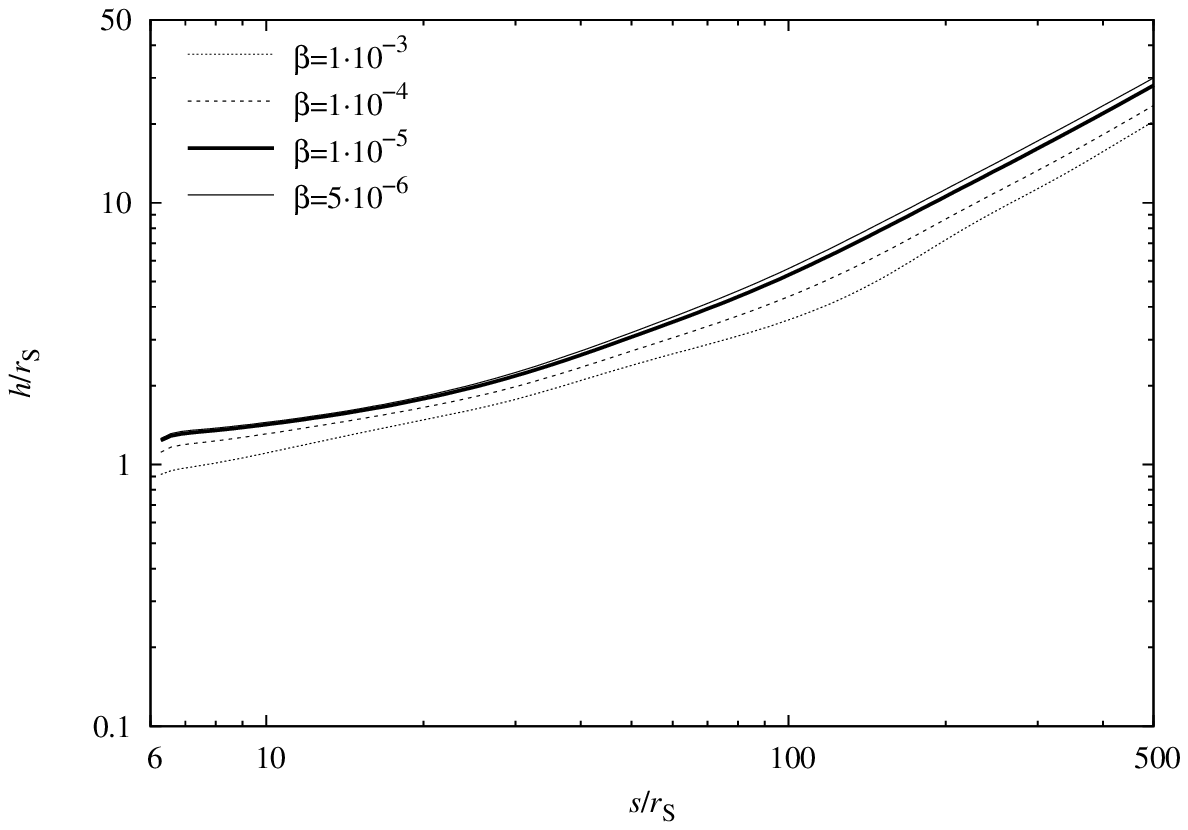}%
\end{minipage}\hfill%
\begin{minipage}{0.47\textwidth}%
\includegraphflex[clip,width=\textwidth]{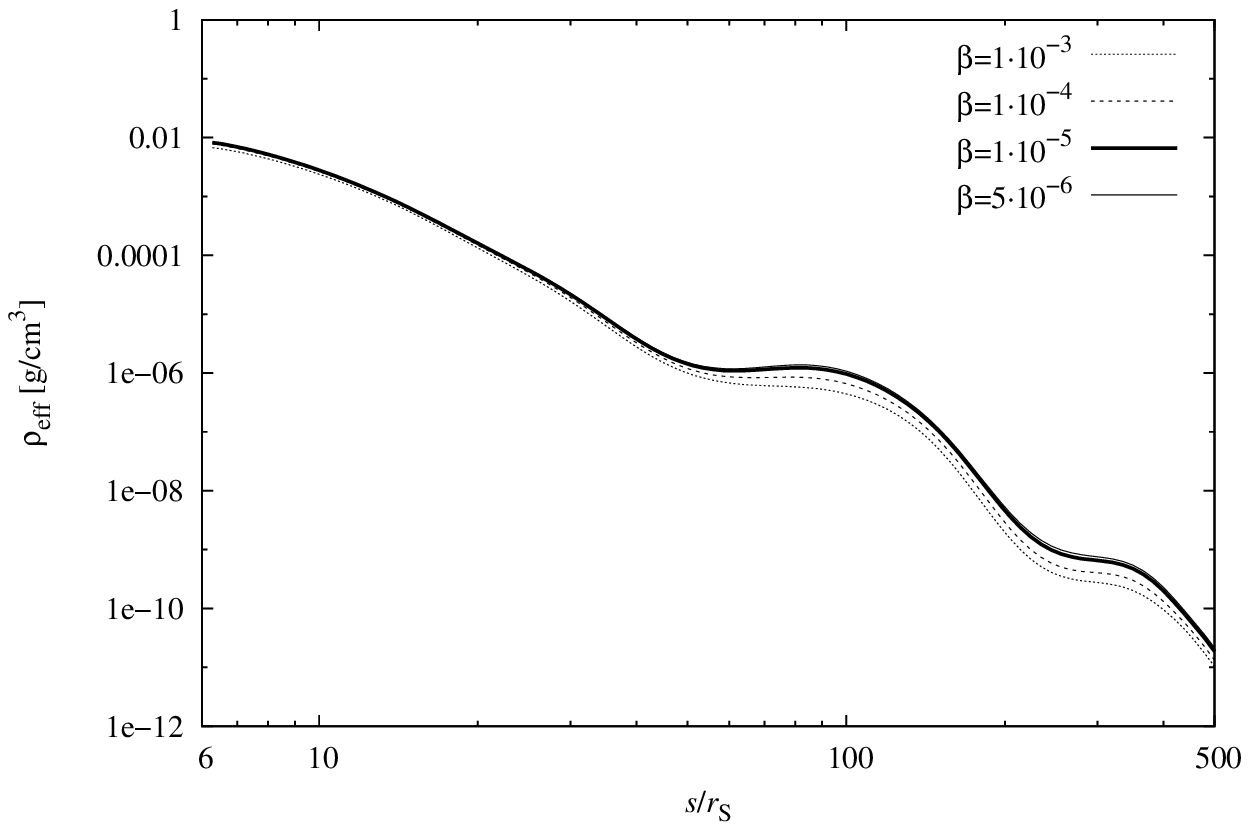}%
\end{minipage}\\%
\vskip-4ex{\small\textbf{(e)}}\rule{0.48\textwidth}{0mm}{\small\textbf{(f)}}\hfill{\small \phantom{x}}\\[1ex]%
\centerline{\begin{minipage}{0.47\textwidth}%
\includegraphflex[clip,width=\textwidth]{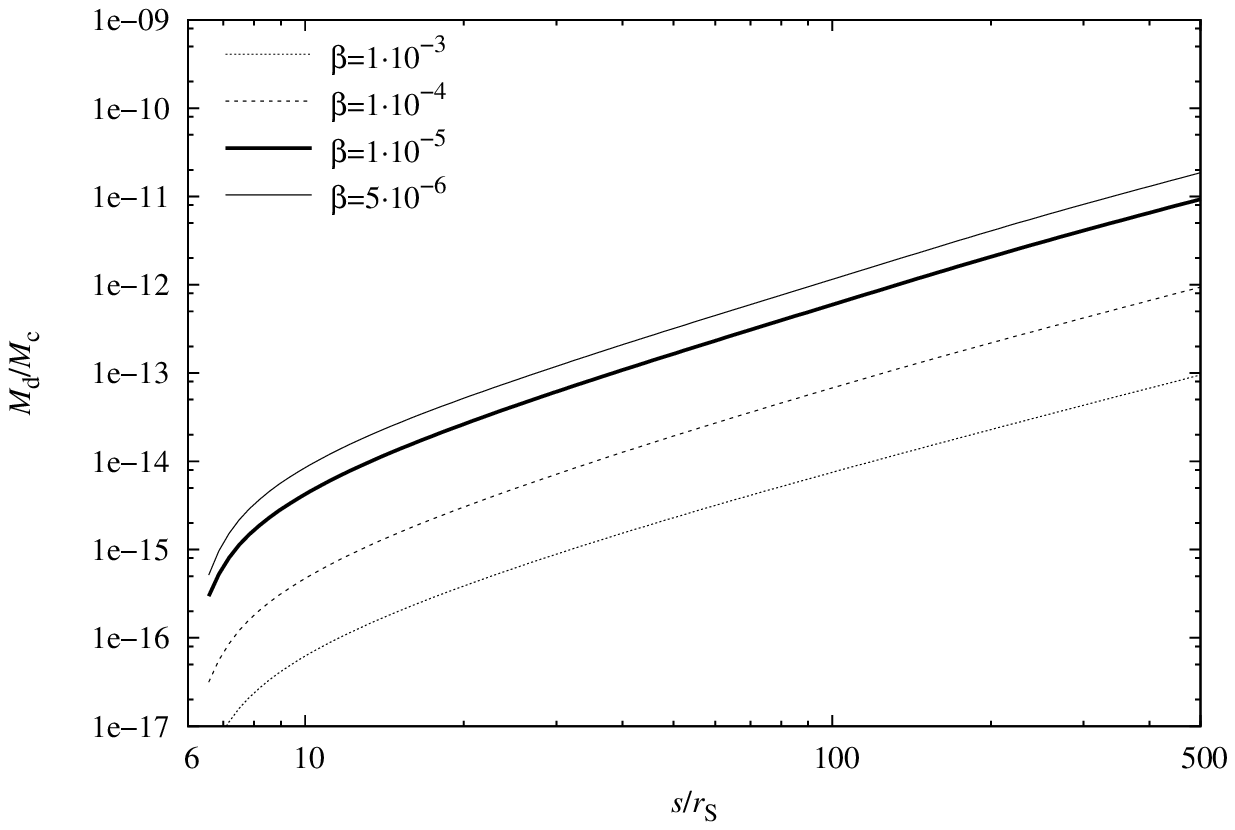}%
\end{minipage}}%
\vskip-4ex\centerline{\begin{minipage}{0.47\textwidth}%
{\small\textbf{(g)}}\hfill\phantom{x}\end{minipage}}%
\caption{Solutions for the standard disc with $m = M_\textup{c}/M_\odot=10$ and $\dot{m}=\dot{M}/\dot{M}_\textup{E}=0.10$ for $\beta=[5\cdot10^{-6};10^{-3}]$}\label{fig_results_std_01}
\end{figure*}

Our main purpose is to investigate the contribution and efficiency of convection in transporting energy and providing viscosity. We therefore
use a standard setup with $M_\textup{c}=10 M_\odot$ and $\dot{M}=0.1 \dot{M}_\textup{E}$ for which we vary the $\beta$-parameter of the
underlying $\beta$-viscosity ($\dot{M}_\textup{E}= 1.67 \cdot 10^{18} \textup{g}/\textup{s}\, (M_\textup{c}/M_\odot)$). We plot the radial structure of these discs in Fig.~\ref{fig_results_std_01} as
a function of radius in units of the Schwarzschild radius $r_\textup{S} = 2.95 \cdot 10^{6}\textup{cm}$.

Common values for the viscosity parameter $\beta$ are in the range of $10^{-4} \ldots 10^{-2}$~\citep{duschl_2000}. To investigate
whether the turbulence caused by convection can account partly for the total viscosity, we perform disc calculations with $\beta = [5\cdot10^{-6};10^{-3}]$.
We limit $\beta$ to this range for the following two reasons.
\begin{enumerate}
\item For $\beta > 10^{-4}$, the standard $\beta$-viscosity prescription causes the turbulent velocity
$v_{\textup{turb},\beta} = \sqrt{\beta} s \omega$ to exceed the sound speed $c_\textup{s}$. In that case, a diffusion limit would have to be introduced \citep{duschl_2000},
resulting in an effective decrease of $\beta$ \citep[see][for a discussion]{heinzeller_2008}. In the particular example of a $10 M_\odot$ black hole accreting
at $10\%$ of the Eddington rate, the diffusion limit sets in for $\beta > 1.4 \cdot 10^{-4}$.
\item For $\beta < 10^{-5}$, hardly any solutions can be found for the required accuracies and the radial range considered here. The reasons therefore will
be revealed hereinafter.
\end{enumerate}
In Fig.~\ref{fig_results_std_01}a, we display the efficiency of convection in the energy transport, measured by the dimensionless quantity $\zeta$.
At each radial position, $\zeta=\zeta(s,z)$ is averaged vertically by
\[
\zeta_\textup{avg} = h^{-1} \int_0^h \zeta dz\,.
\]
For $\beta \geq 10^{-4}$, we find that a significant amount of the total energy is transported by convection in the inner part of the disc;
close to the inner disc radius, $\zeta \approx 0.98$ for $\beta=10^{-3}$. Radiative energy transport dominates in the outer part of these discs, with a transition
zone expanding from $[10 r_\textup{S}; 80 r_\textup{S}]$ for $\beta=10^{-3}$ to $[10 r_\textup{S}; 180 r_\textup{S}]$ for smaller $\beta$. While
the curves show a smooth behavior for $\beta \geq 10^{-5}$, this picture changes when $\beta$ is decreased further. Radial variations of $\zeta$ of about
$0.1$ occur in the case $\beta = 5\cdot10^{-6}$, for which we also plot a fitting curve. In general, smaller supporting viscosities (i.\,e., smaller values
of $\beta$) have little influence on the outer regions, while they lead to a significant decrease of the efficiency of convective energy transport in the
inner disc region.

For a proper explanation of the possible reasons for these variations in $\zeta$, we display further disc quantities in Figs.~\ref{fig_results_std_01}b--g.
The relative contribution of the radiation pressure to the total pressure in the disc's mid plane is shown in Fig.~\ref{fig_results_std_01}b. A comparison
with the efficiency of convection, described by the quantity $\zeta$, nicely confirms theoretical expectations that a strong radiation pressure inside the disc drives the
convective motion -- a simple linear correlation, however, cannot be found. We would like to point out that both the gas and the radiation pressure do not
reflect the instabilities in $\zeta$.

To examine the influence of convective turbulence on the disc viscosity, we further display
$\beta_\textup{avg} = \nu_\textup{avg}/\nu_\beta$, where $\nu_\beta$ is constant for the vertical stratification and 
\[
\nu_\textup{avg} = h^{-1} \int_{0}^{h} \nu\,dz = h^{-1} \int_{0}^{h} (\nu_\beta +\nu_\textup{conv})\,dz\,.
\]
In the low-$\beta$ case, the convective viscosity $\nu_\textup{conv}$ becomes three times as large as the
underlying $\beta$-viscosity. It is important to note that although the convective viscosity becomes \emph{relatively stronger} for lower supporting
viscosities, its absolute value decreases as well. As before, instabilities occur for $\beta < 10^{-5}$, which are displayed together with the corresponding
fitting curve.

The surface density $\Sigma$ increases almost linearly with $\beta$ and reflects the variations of $\zeta$ only very weakly. Since the disc scale height $h$ and the
density at the disc surface $\rho_\textup{eff}$ are both almost unaffected by the value of $\beta$, the increase in $\Sigma$ is due to a larger internal density in the disc.
In all cases, the discs are geometrically thin in the outer part, and ``slim'' in the inner part, with a maximum ratio of $h/s\approx 0.2$. Contrary to the case of $\zeta$,
no instabilities are found in $h$ for the low-$\beta$ case. The density $\rho_\textup{eff}$ at the disc's surface shows very similar results for all solutions with a clear decreasing trend towards larger radii.
A certain irregular structure can be seen for all results, an effect of the opacity model, which itself is very sensitive to the densities and temperatures
in this region of the disc. As for the pressure and the disc height, the instabilities in $\zeta$ are not reflected in the density.

The disc mass $M_\textup{d}$ increases for decreasing $\beta$, but remains completely negligible for all models.
We estimate the equality radius $s_\textup{equ}$ where $M_\textup{d}(s)=M_\textup{c}$ by extrapolating the results towards larger radii for the $\beta=10^{-5}$ disc case.
A linear fit to the outer region in the $\log$-$\log$ plot gives
\begin{equation}
M_\textup{d}(s)/M_\textup{c} = 1\cdot 10^{-11} \left(\frac{s}{500 r_\textup{S}}\right)^{1.75}\!\!\!,\quad  \beta=5 \cdot 10^{-6}\,,\label{eqn_scaling_law_Mdisc}
\end{equation}
which in turn leads to $s_\textup{equ} = 6.7 \cdot 10^{7} r_\textup{S}$. Thus, self-gravity is safely negligible in our disc calculations. By means of the
radial disc equations~\eqref{eqn_continuity_equation}--\eqref {eqn_energy}, this implies the same results for the total heat flux $F$ and therefore
for the temperature $T_\textup{eff}$ at the disc surface (not shown here), regardless of the value of $\beta$. Furthermore, this also implies that the
radial variations of $\zeta$ have no effect on the temperature profile. In summary, the irregularities of the efficiency of convective energy transport are
reflected weakly in $\Sigma$, but have no influence on the remaining physical quantities.
\subsection{Extended parameter space} 
In this section, we extend the disc calculations towards varying accretion rates and central masses in order to
see how general properties and, in particular, the instabilities in $\zeta$, depend on the input parameters.
\subsubsection{Eddington ratio $\dot{M}/\dot{M}_\textup{E}$}\label{sec_accretion_rate}
\begin{figure*}%
\begin{minipage}{0.47\textwidth}%
\includegraphflex[clip,width=\textwidth]{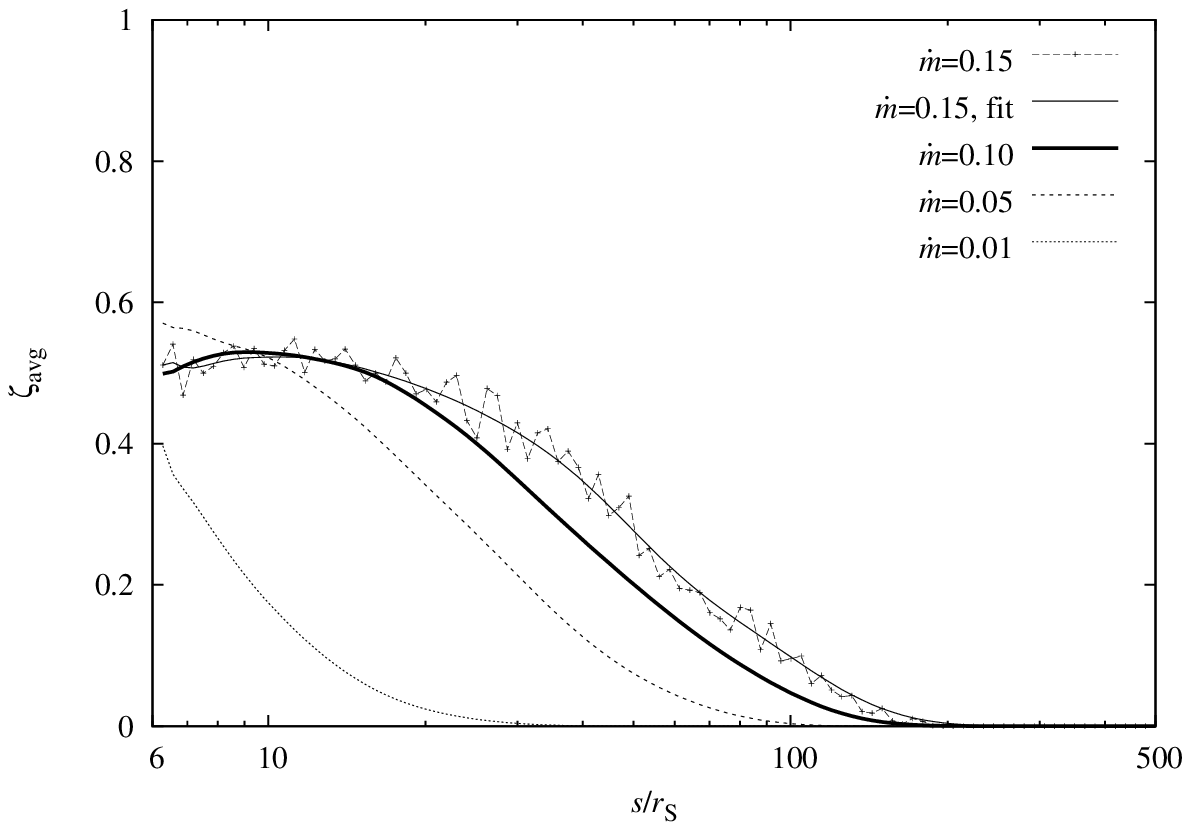}%
\end{minipage}\hfill%
\begin{minipage}{0.47\textwidth}%
\includegraphflex[clip,width=\textwidth]{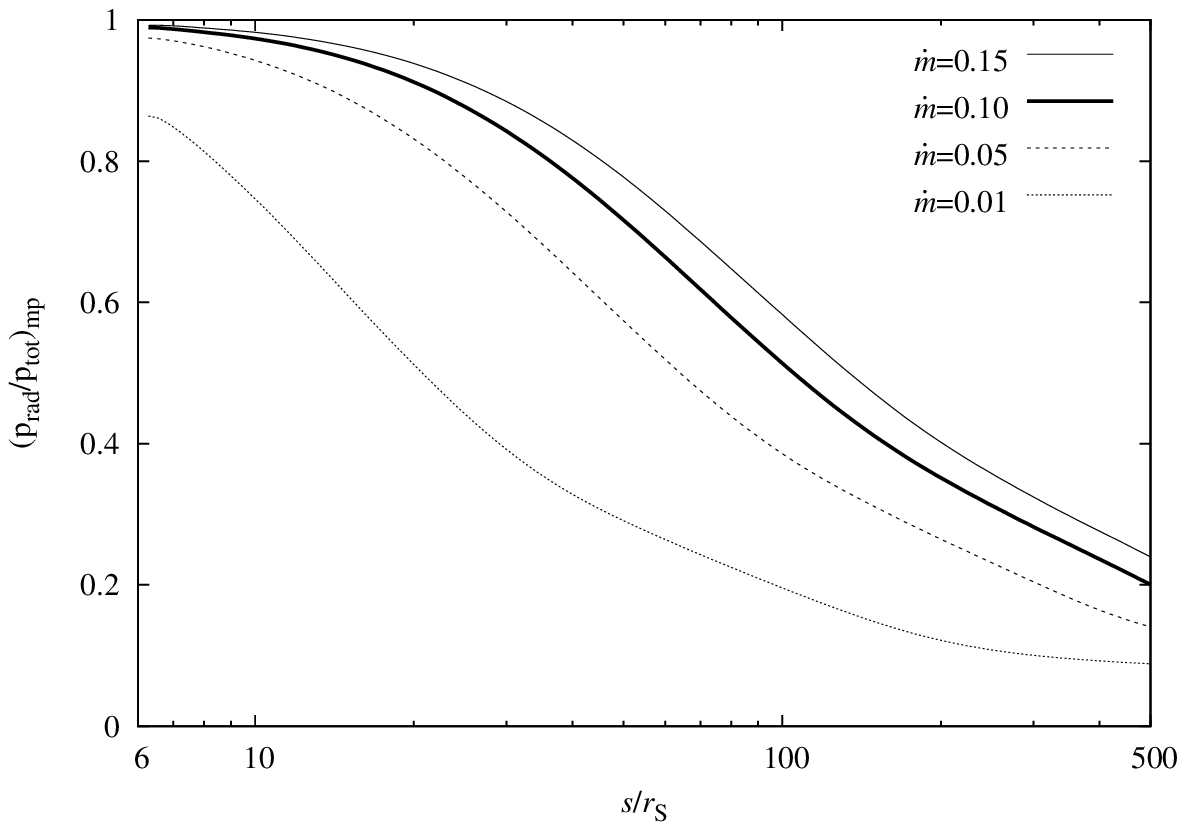}%
\end{minipage}\\%
\vskip-4ex{\small\textbf{(a)}}\rule{0.48\textwidth}{0mm}{\small\textbf{(b)}}\hfill{\small \phantom{x}}\\[1ex]%
\begin{minipage}{0.47\textwidth}%
\includegraphflex[clip,width=\textwidth]{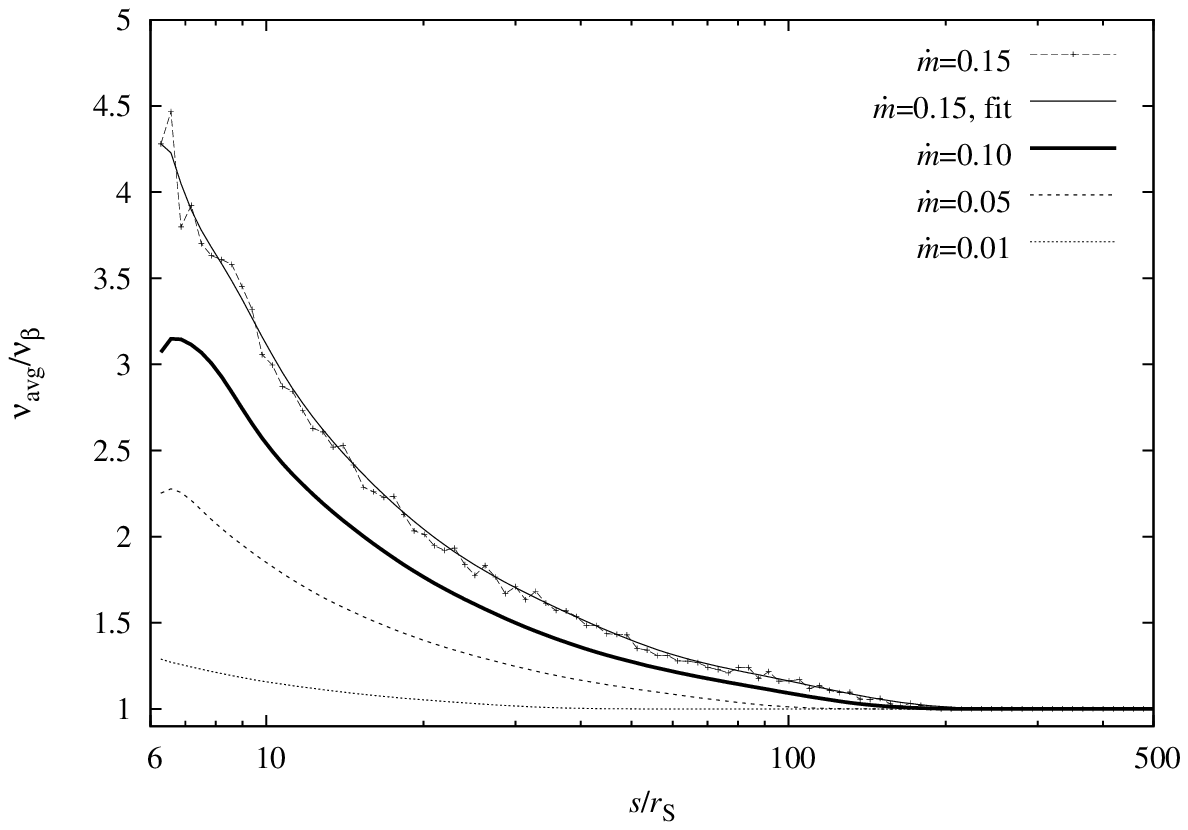}%
\end{minipage}\hfill%
\begin{minipage}{0.47\textwidth}%
\includegraphflex[clip,width=\textwidth]{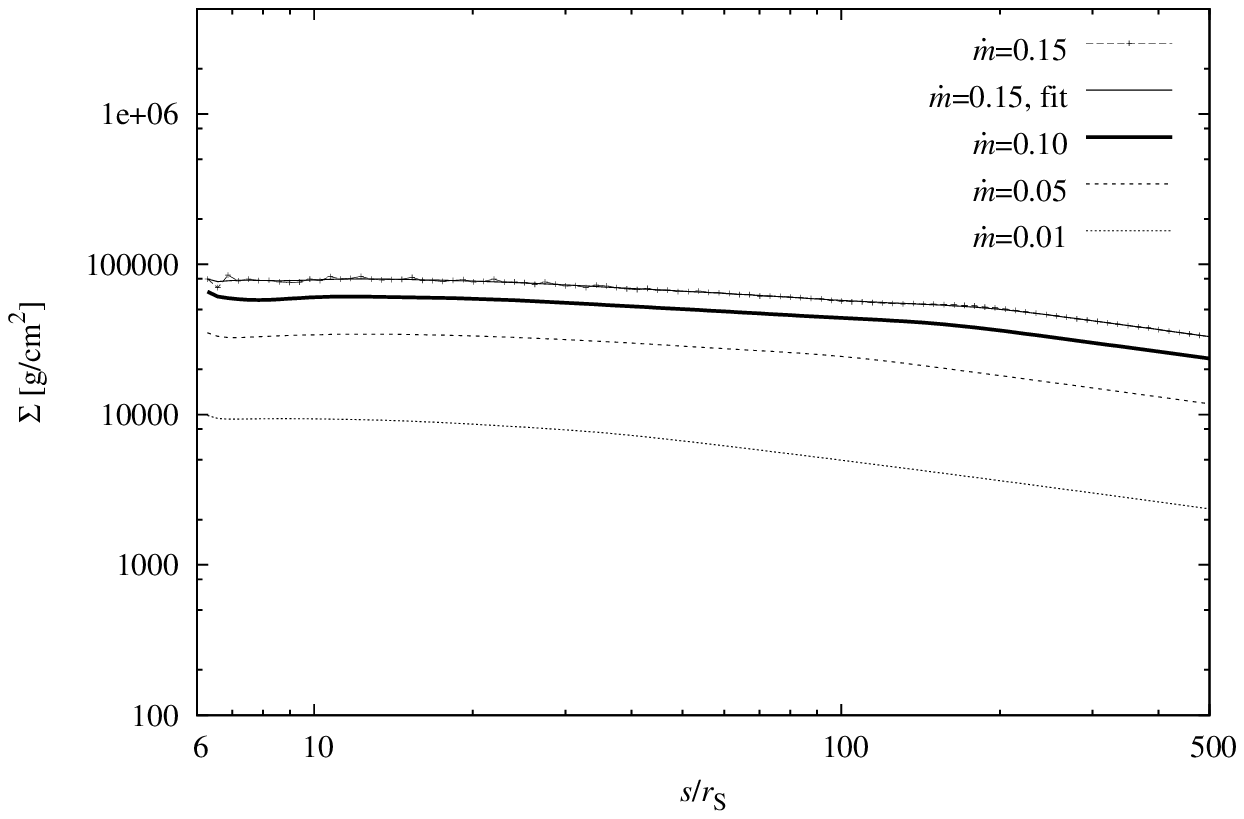}%
\end{minipage}\\%
\vskip-4ex{\small\textbf{(c)}}\rule{0.48\textwidth}{0mm}{\small\textbf{(d)}}\hfill{\small \phantom{x}}\\[1ex]%
\begin{minipage}{0.47\textwidth}%
\includegraphflex[clip,width=\textwidth]{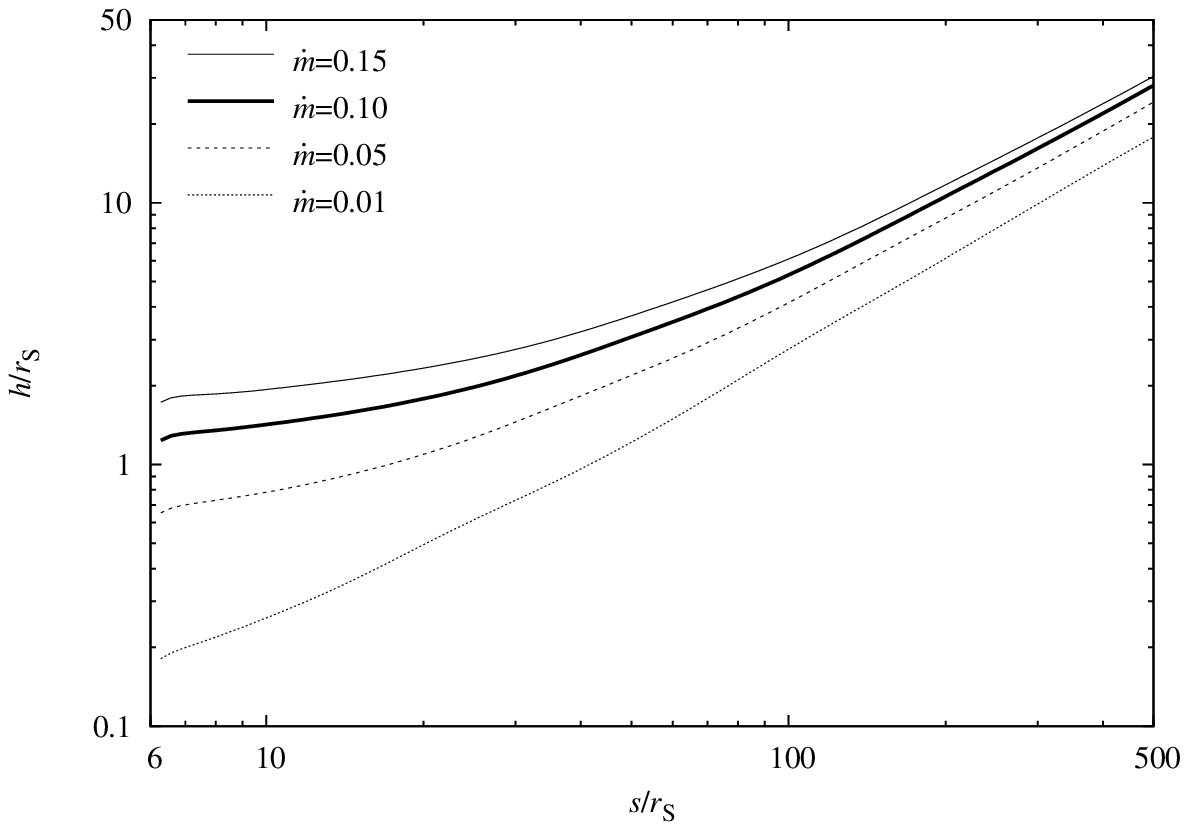}%
\end{minipage}\hfill%
\begin{minipage}{0.47\textwidth}%
\includegraphflex[clip,width=\textwidth]{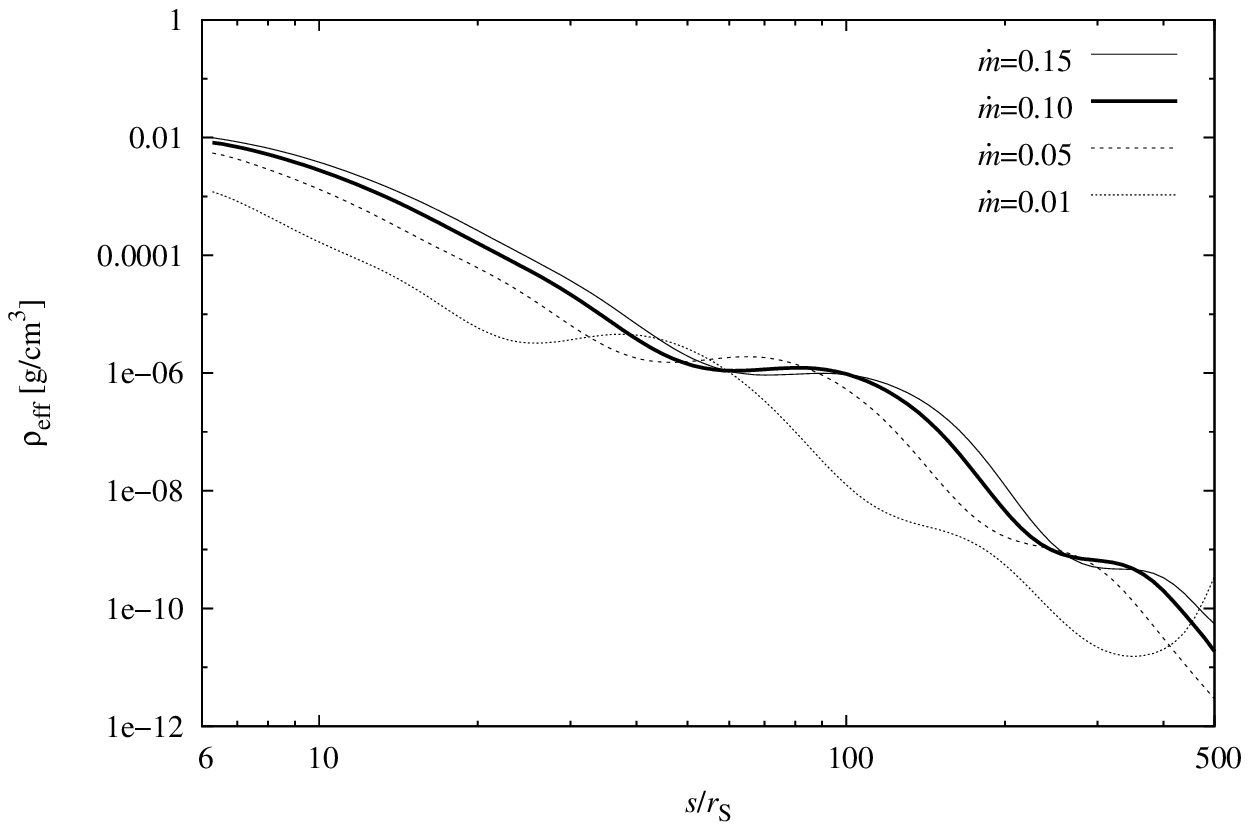}%
\end{minipage}%
\vskip-4ex{\small\textbf{(e)}}\rule{0.48\textwidth}{0mm}{\small\textbf{(f)}}\hfill{\small \phantom{x}}\\[1ex]%
\centerline{\begin{minipage}{0.47\textwidth}%
\includegraphflex[clip,width=\textwidth]{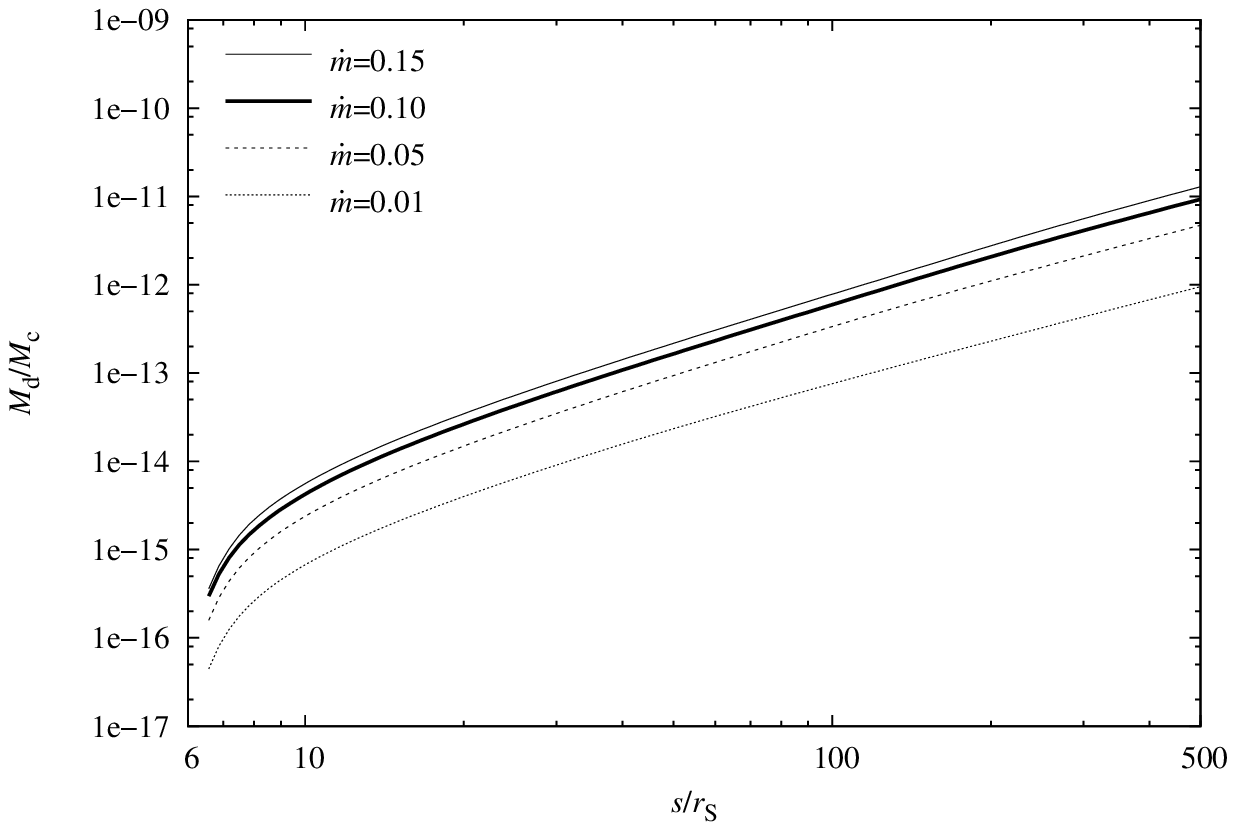}%
\end{minipage}}%
\vskip-4ex\centerline{\begin{minipage}{0.47\textwidth}%
{\small\textbf{(g)}}\hfill\phantom{x}\end{minipage}}%
\caption{Solutions for varying accretion rates $\dot{m}=\dot{M}/\dot{M}_\textup{E}$ with $m=M_\textup{c}/M_\odot=10$ and $\beta=10^{-5}$}\label{fig_Mdotvar_final_01}
\end{figure*}
In the first step, we investigate the dependence of the results on the accretion rate while keeping a constant $\beta=10^{-5}$ and a constant
$M_\textup{c}=10 M_\odot$. We perform disc calculations with accretion rates of $\dot{M}=[0.01;0.15] \dot{M}_\textup{E}$. Higher rates
are not included, since the discs become too thick for the thin-disc approximation to be valid: for $\dot{M}=0.15 \dot{M}_\textup{E}$, the ratio $h/s$ reaches values
of $0.3$ in the inner disc region, while it does not exceed $0.03$ for the lower limit $\dot{M}=0.01$ (see Fig.~\ref{fig_Mdotvar_final_01}).
Furthermore, the same type of radial variations in $\zeta$ occur for $\dot{M} > 0.1 \dot{M}_\textup{E}$, which prevent the computations to converge for higher values
of the accretion rate. For illustration, we display them along with the fitting curve in Fig.~\ref{fig_Mdotvar_final_01}a.
{For the stable solutions ($\dot{M} \leq 0.1 \dot{M}_\textup{E}$), the relative contribution of convection to the overall
energy transport is smaller for lower accretion rates. This is because both} the total energy and the angular momentum that have to be transported through the
disc depend linearly on the accretion rate (c.\,f.,~\eqref{eqn_energy},~\eqref{eqn_alpha_disc}). Thus, the standard $\beta$-viscosity is almost
large enough to account for both requirements when the accretion rate is low.
{Higher accretion rates than $0.1 \dot{M}_\textup{E}$} lead to the
same type of instabilities of $\zeta$ as lower $\beta$-values $< 10^{-5}$ do for the standard disc setup ($M_\textup{c}=10M_\odot$, $\dot{M}=0.1\dot{M}_\textup{E}$).
{Like} in the previous section, a higher convective efficiency corresponds to a higher contribution of the radiation pressure to the total pressure.

Figure~\ref{fig_Mdotvar_final_01} further demonstrates that the disc mass and the surface density scale almost linearly with the
accretion rate; opacity effects modify this scaling law in case of the density $\rho_\textup{eff}$ at the disc surface.
The contribution of convective turbulence is naturally higher the higher the accretion rate is, up to $\nu_\textup{conv}= 3 \nu_\beta$.
The convective zone reaches outwards to $20r_\textup{S}$ for low accretion rates, and to $200r_\textup{S}$ for high accretion rates, respectively.

We want to note that the self-gravity of the disc remains negligible and therefore the effective temperature scales with
$T_\textup{eff} \propto \dot{M}^{0.25}$, as expected from the radial structure equations. As in the previous case, the strong
variations in $\zeta$ are reflected only in the viscosity $\nu_\textup{avg}$ and the surface density $\Sigma$, though rather weakly.
\subsubsection{Central black hole mass $M_\textup{c}$}\label{sec_central_mass}
\begin{figure*}%
\begin{minipage}{0.47\textwidth}%
\includegraphflex[clip,width=\textwidth]{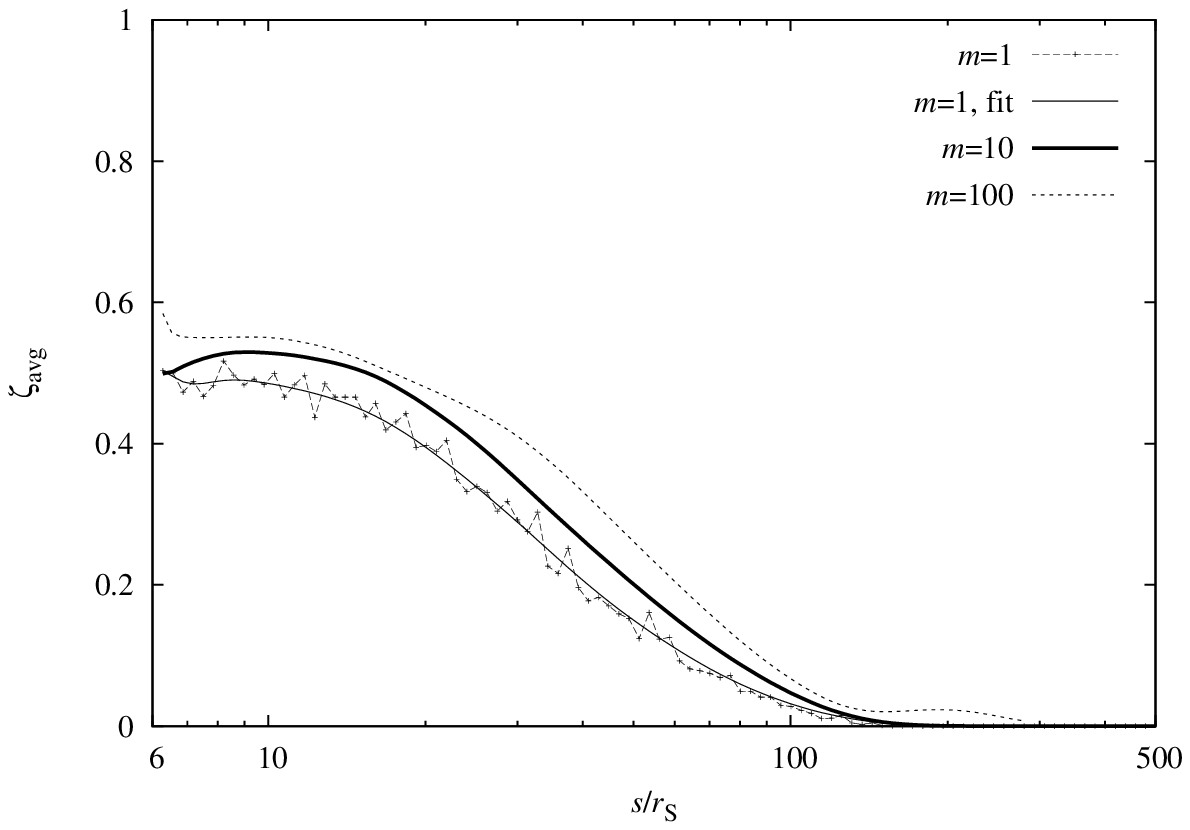}%
\end{minipage}\hfill%
\begin{minipage}{0.47\textwidth}%
\includegraphflex[clip,width=\textwidth]{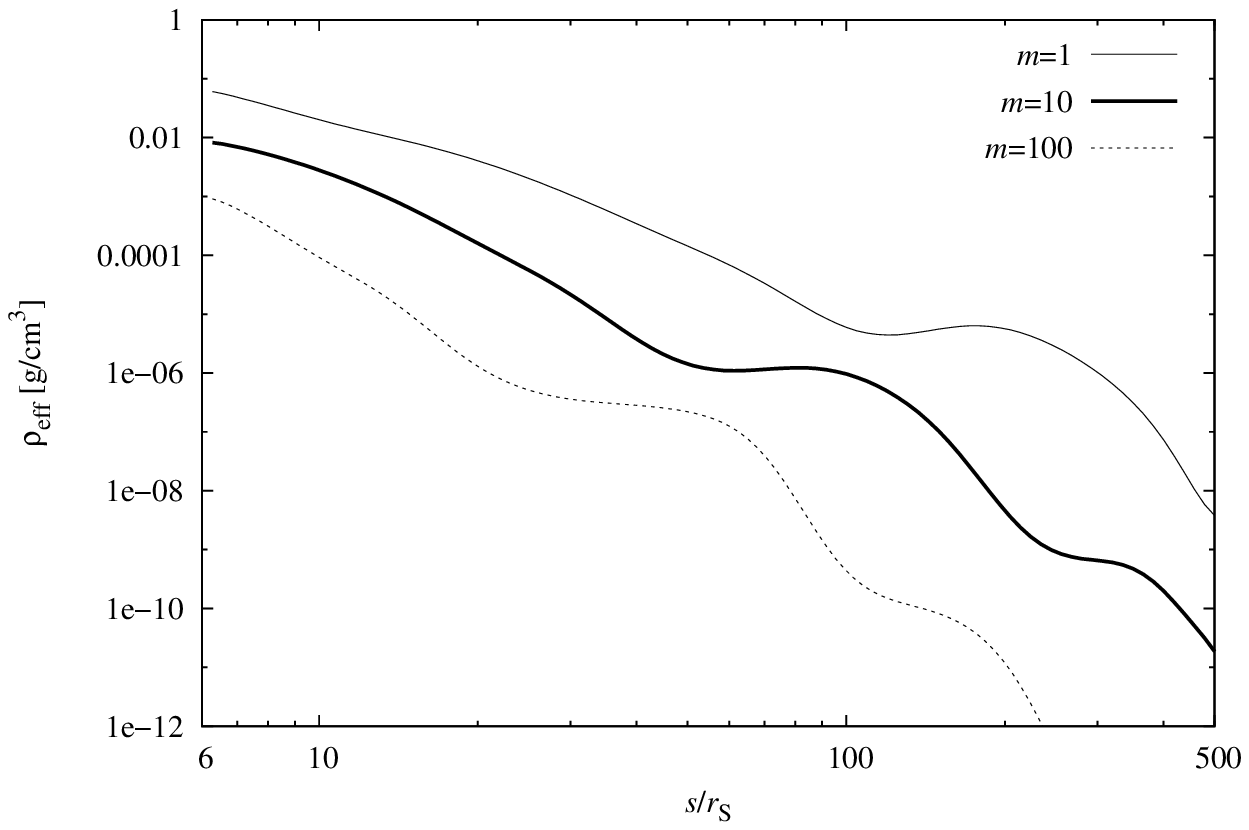}%
\end{minipage}\\%
\vskip-4ex{\small\textbf{(a)}}\rule{0.48\textwidth}{0mm}{\small\textbf{(b)}}\hfill{\small \phantom{x}}\\[1ex]%
\begin{minipage}{0.47\textwidth}%
\includegraphflex[clip,width=\textwidth]{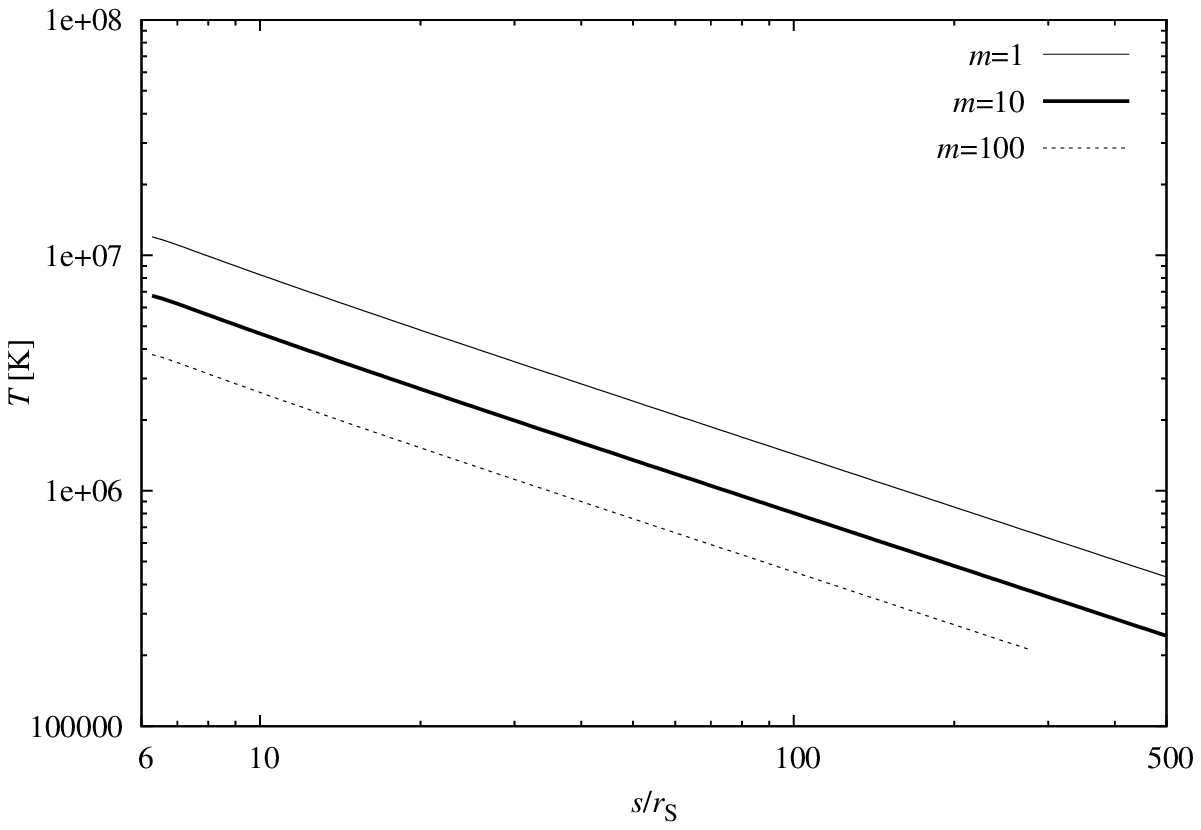}%
\end{minipage}\hfill%
\begin{minipage}{0.47\textwidth}%
\includegraphflex[clip,width=\textwidth]{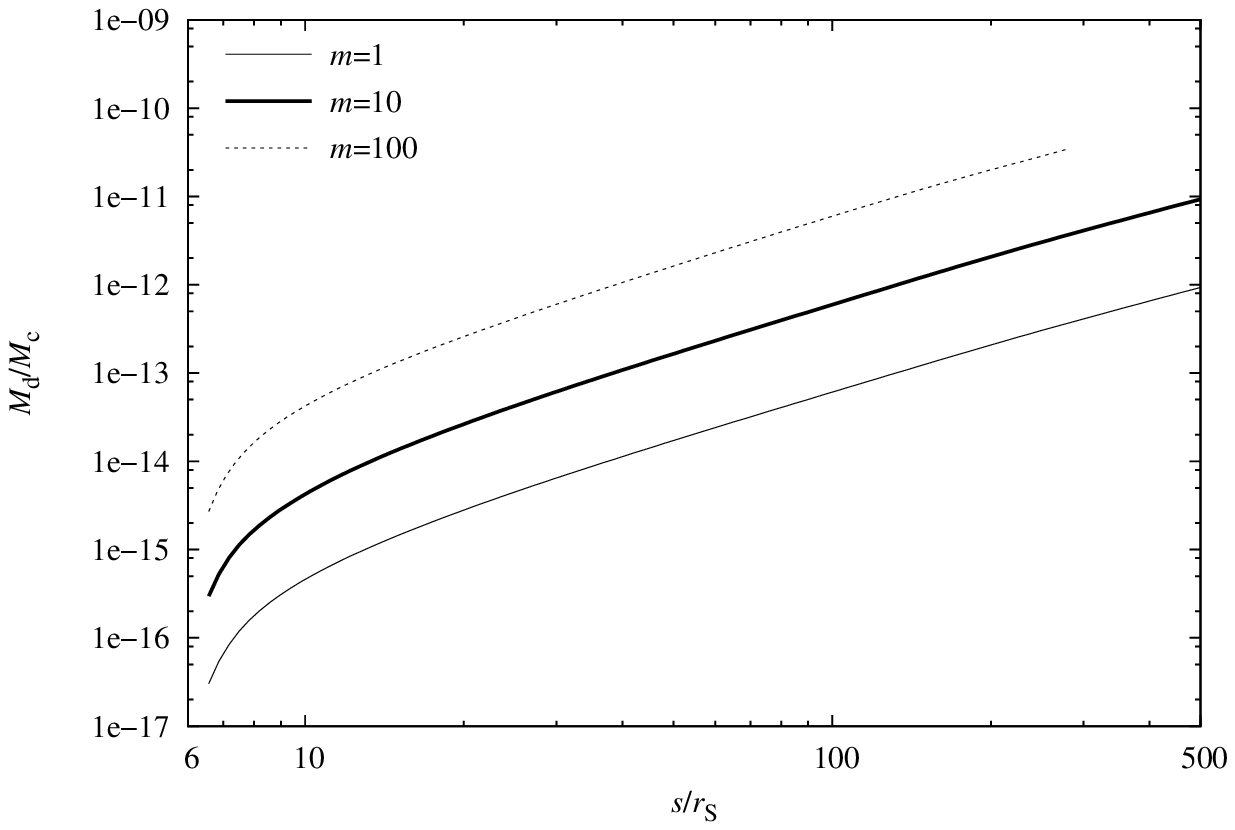}%
\end{minipage}%
\vskip-4ex{\small\textbf{(c)}}\rule{0.48\textwidth}{0mm}{\small\textbf{(d)}}\hfill{\small \phantom{x}}%
\caption{Solutions for varying central masses $m=M_\textup{c}/M_\odot$, constant Eddington ratio $\dot{m}=\dot{M}/\dot{M}_\textup{E} = 0.10$
and $\beta=10^{-5}$}\label{fig_Mcenvar_final_01}
\end{figure*}
Next, we vary the mass of the central black hole in the range of $M_\textup{c}=[1;100] M_\odot$ while keeping a constant
$\beta=10^{-5}$ and a constant Eddington ratio $\dot{M} = 0.1 \dot{M}_\textup{E}$ for the accretion rate. \emph{Thus, the absolute
value of the accretion rate is implicitly scaled with $M_\textup{c}$}. Figure~\ref{fig_Mcenvar_final_01}
displays selected properties of these disc models. The efficiency of convection $\zeta$ is almost insensitive on a varying central mass,
with the small differences being to due the disc temperature (Figs.~\ref{fig_Mcenvar_final_01}a,c). Depending on the dissipation rate
solely, the effective temperature scales with $M_\textup{c}^{-0.25}$ (see~\eqref{eqn_energy}). At the same time, $\nu_\beta \propto M_\textup{c}$ (see \eqref{eqn_nu_beta}).
Thus, less energy has to be transported through the vertical layers, while at the same time the supporting viscosity is increased for higher central masses. 
Radial variations in $\zeta$ set in for the lowest central mass case, $M_\textup{c}=1M_\odot$. For completion, we would
like to add that the ratio $\nu_\textup{avg}/\nu_\beta$ is almost independent of the central mass when the Eddington ratio is kept constant, in accordance to the
behavior of $\zeta$.

It turns out that both the ratio $h/r_\textup{S}$ and the surface density $\Sigma$ do not change for varying central masses.
For this fact to hold, the mass density $\rho$ has to scale with $M_\textup{c}^{-1}$, which is reflected nicely in Fig.~\ref{fig_Mcenvar_final_01}b.
Then, given that the disc's mass depends only on $s^2 \propto M_\textup{c}^2$, the ratio $M_\textup{d}/M_\textup{c}$ scales with $M_\textup{c}$ (Fig.~\ref{fig_Mcenvar_final_01}d).
Note that the calculation for the $M_\textup{c}= 100 M_\odot$ case terminates at $s\approx 300 r_\textup{S}$,
since the density $\rho_\textup{eff}$ decreases to $10^{-12} \textup{g}/\textup{cm}^3$, which is the value of
the upper boundary condition $\rho_\textup{up}$ in the atmosphere.

Since the gas pressure ($\propto M_\textup{c}^{-1.25}$) decreases more rapidly than the radiation pressure ($\propto M_\textup{c}^{-1}$)
with increasing central mass, the ratio $p_\textup{rad}/p_\textup{tot}$ is higher the larger the central mass.
\subsubsection{Central black hole mass and Eddington ratio}\label{sec_central_mass_and_accretion_rate}
Finally, we investigate the case of varying central masses for a constant $\beta=10^{-5}$ and a constant absolute value $\dot{M}$ such that
it equals $0.1\dot{M}_\textup{E}$ for a $10 M_\odot$ black hole. \emph{Thus, the Eddington ratio $\dot{M}/\dot{M}_\textup{E}$
scales with $M_\textup{c}^{-1}$}\!. We consider a parameter range of $M_\textup{c}=[6.7;100] M_\odot$, corresponding to Eddington ratios
of $\dot{M} = [0.15;0.01]\dot{M}_\textup{E}$. For even lower central masses (i.\,e., higher Eddington ratios), the discs get too thick
and also the radial variations of $\zeta$ become to pronounced to let the calculations converge.
\begin{figure*}%
\begin{minipage}{0.47\textwidth}%
\includegraphflex[clip,width=\textwidth]{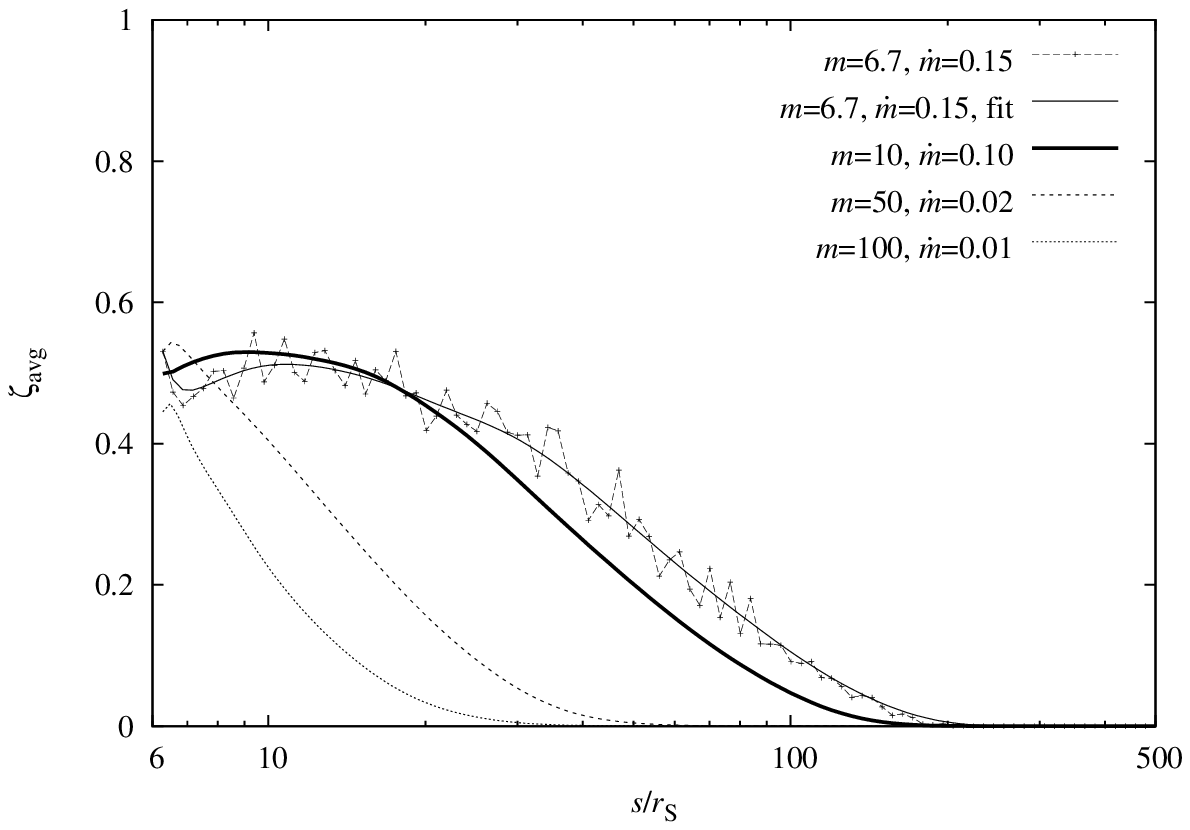}%
\end{minipage}\hfill%
\begin{minipage}{0.47\textwidth}%
\includegraphflex[clip,width=\textwidth]{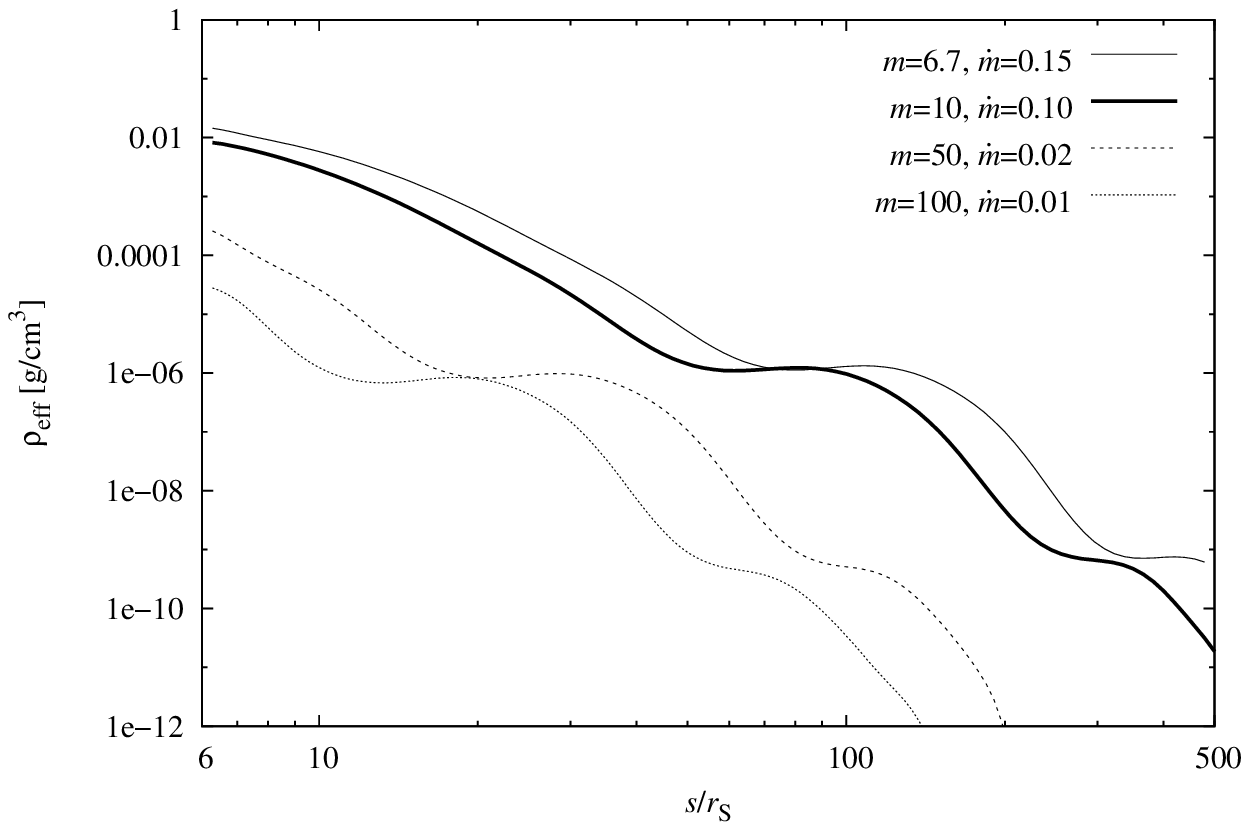}%
\end{minipage}\\%
\vskip-4ex{\small\textbf{(a)}}\rule{0.48\textwidth}{0mm}{\small\textbf{(b)}}\hfill{\small \phantom{x}}\\[1ex]%
\begin{minipage}{0.47\textwidth}%
\includegraphflex[clip,width=\textwidth]{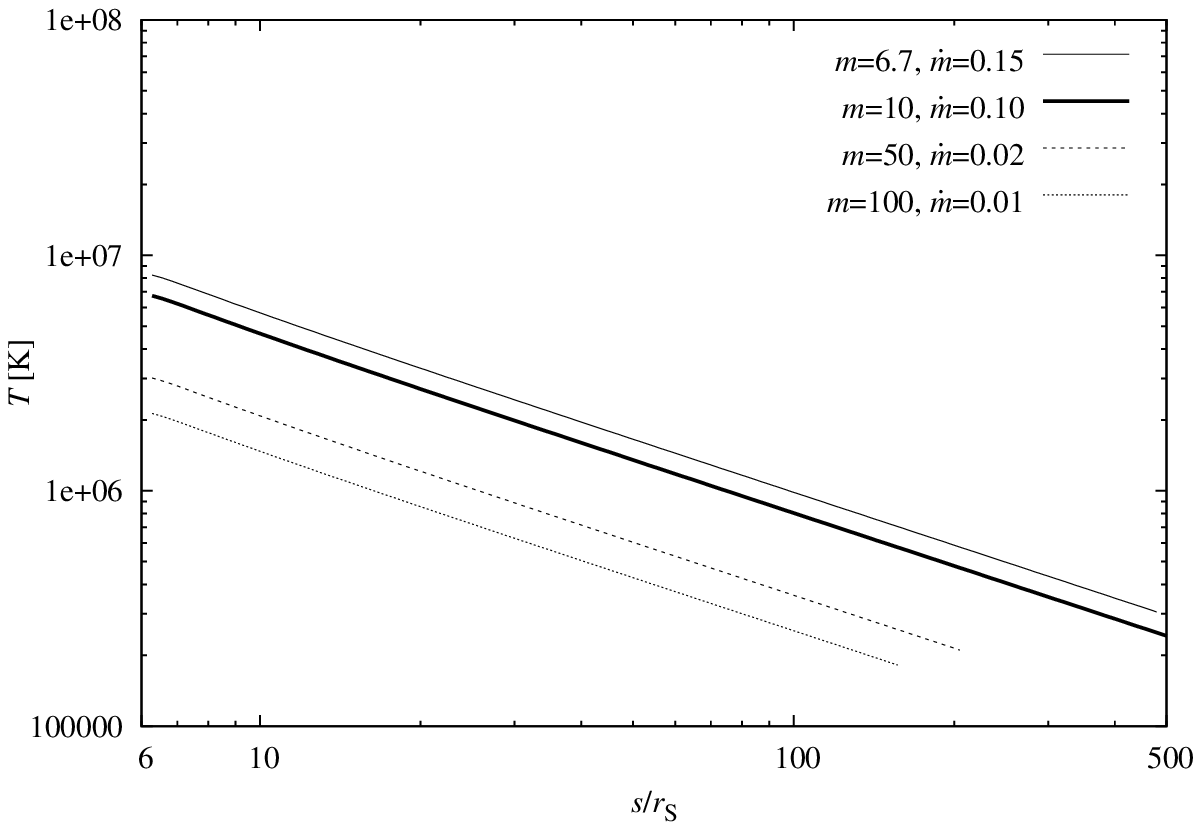}%
\end{minipage}\hfill%
\begin{minipage}{0.47\textwidth}%
\includegraphflex[clip,width=\textwidth]{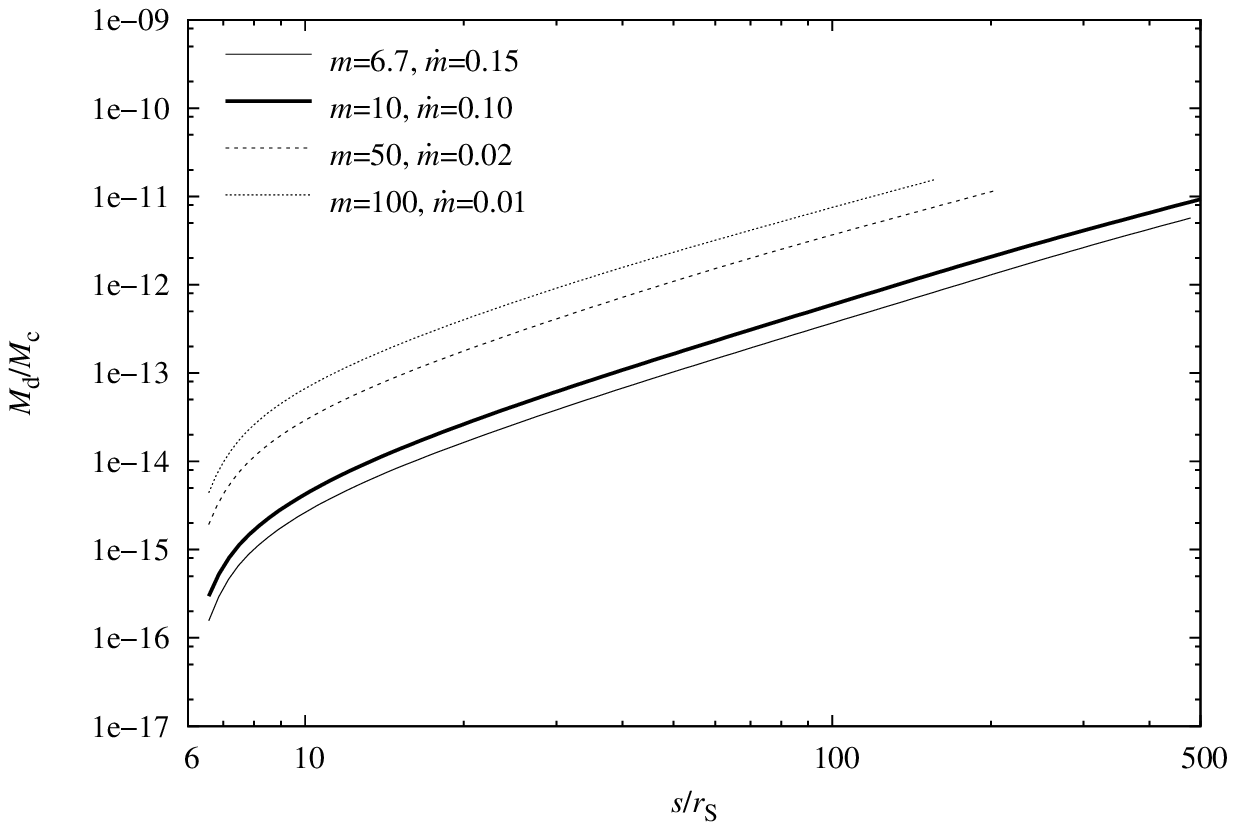}%
\end{minipage}%
\vskip-4ex{\small\textbf{(c)}}\rule{0.48\textwidth}{0mm}{\small\textbf{(d)}}\hfill{\small \phantom{x}}%
\caption{Solutions for varying central masses $m=M_\textup{c}/M_\odot$ with constant $\dot{M}$ -- corresponding to $\dot{m}=\dot{M}/\dot{M}_\textup{E}=0.10$ for $m=10$ -- and $\beta=10^{-5}$}\label{fig_Mcenvar_Mdotvar_final_02}
\end{figure*}

Figure~\ref{fig_Mcenvar_Mdotvar_final_02} displays selected properties of these discs, which are now controlled by the combined effects of a varying Eddington ratio
and central black hole mass. The efficiency of convection, expressed by $\zeta$ in
Fig.~\ref{fig_Mcenvar_Mdotvar_final_02}a, is determined mainly by the Eddington ratio and is thus similar to Sect.~\ref{sec_accretion_rate}.
 In the case of the density at the disc surface $\rho_\textup{eff}$ and the temperature $T_\textup{eff}$ (Figs.~\ref{fig_Mcenvar_Mdotvar_final_02}b,c),
the two effects enforce each other, while the disc mass $M_\textup{d}$ is vastly controlled by the central mass and therefore scales as in Sect.~\ref{sec_central_mass} (Fig.~\ref{fig_Mcenvar_Mdotvar_final_02}d).

The pressure ratio $p_\textup{rad}/p_\textup{tot}$ shows an inverse behavior than in Sect.~\ref{sec_central_mass}, which corresponds to higher convective efficiencies $\zeta$
for lower central masses (i.\,e., higher accretion rates). This inverse behavior is due to the fact that, here, the increase in $p_\textup{rad}\propto T^4$ is stronger than
the increase in $p_\textup{gas}\propto\rho\cdot T$ for lower central masses.
\subsection{Radial variations in the convection efficiency {\boldmath $\zeta$}}
An important point in this discussion is the origin
of the instabilities in $\zeta$ for certain disc solutions. We have seen that they occur if the underlying viscosity is decreased
under a threshold value, which itself depends on the parameters central mass and accretion rate. Interestingly, these instabilities
appear predominantly in $\zeta$ and $\nu_\textup{avg}/\nu_\beta$ and only weakly in the surface density $\Sigma$. The effect
on the remaining physical quantities is negligible or zero, especially for the observables such as the effective temperature.

\begin{figure}%
\includegraphflex[clip,width=\columnwidth]{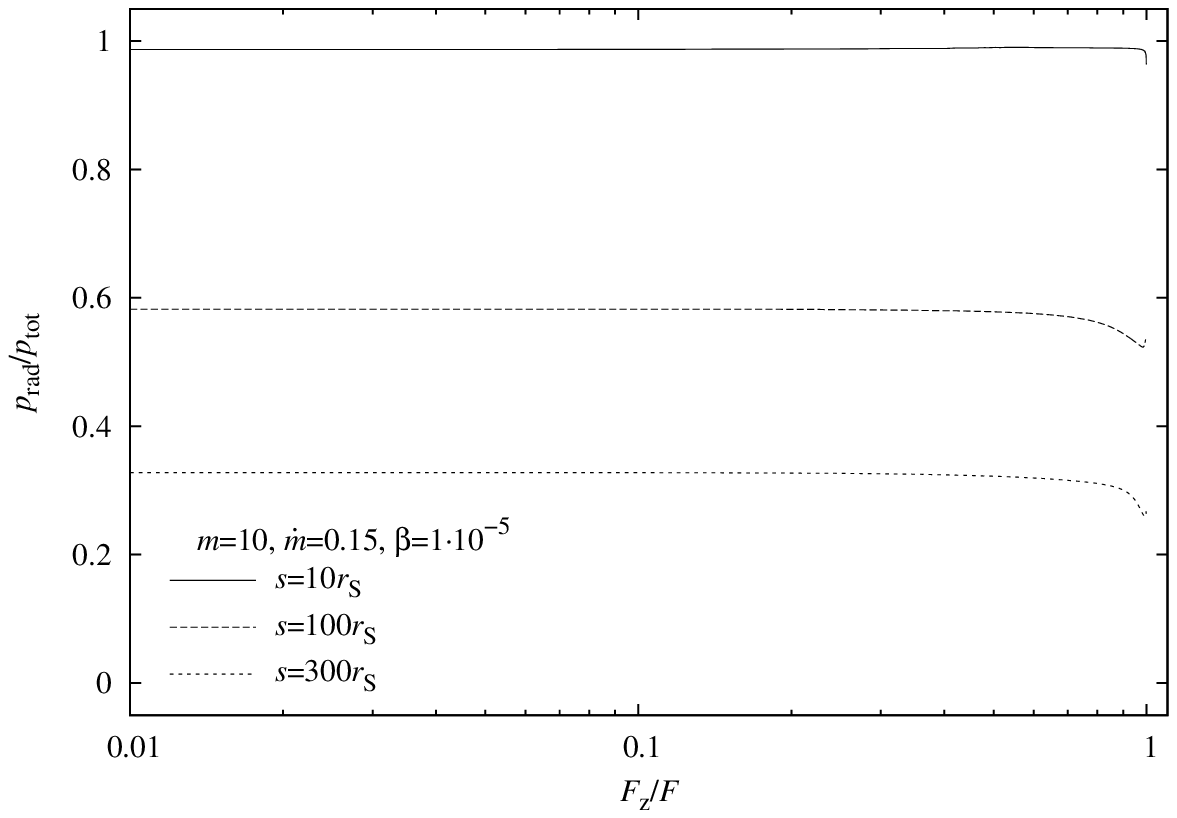}\\%
\vskip-6ex{\small\textbf{(a)}}\\[1ex]%
\includegraphflex[clip,width=\columnwidth]{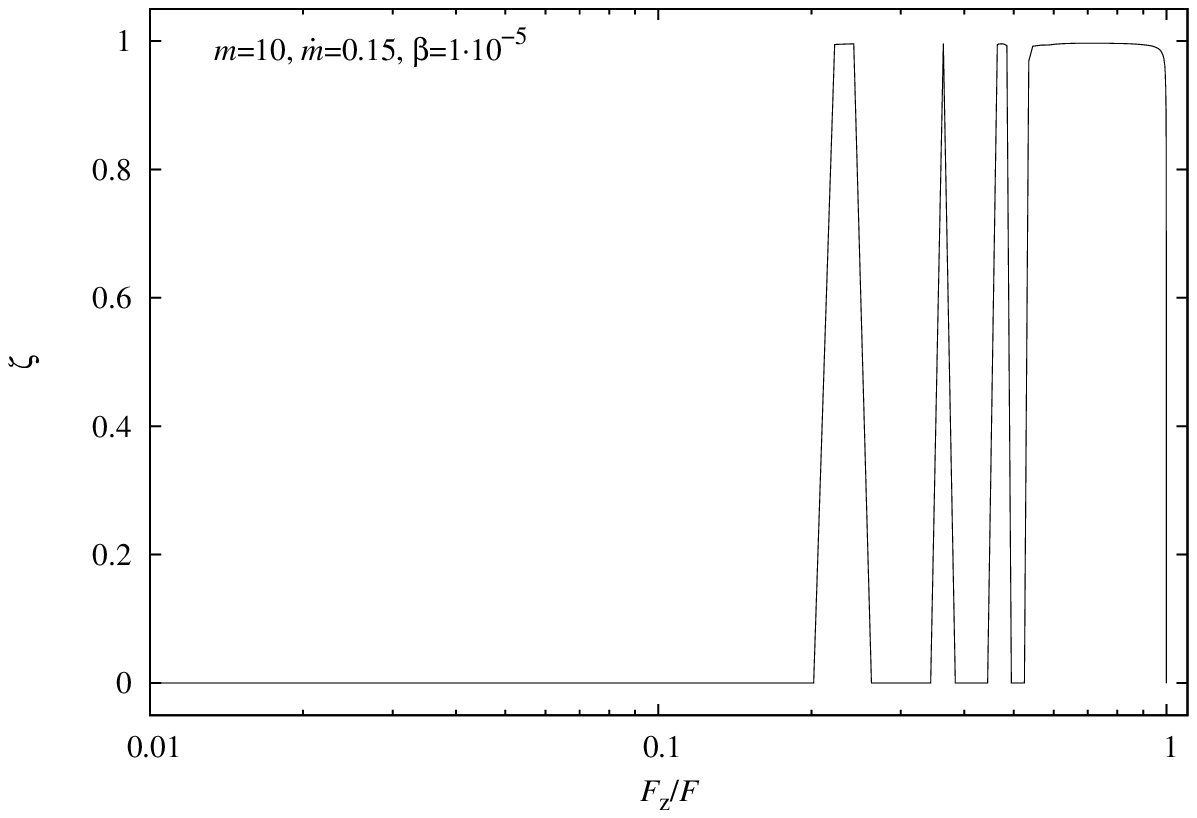}\\%
\vskip-6ex{\small\textbf{(b)}}\\[1ex]%
\includegraphflex[clip,width=\columnwidth]{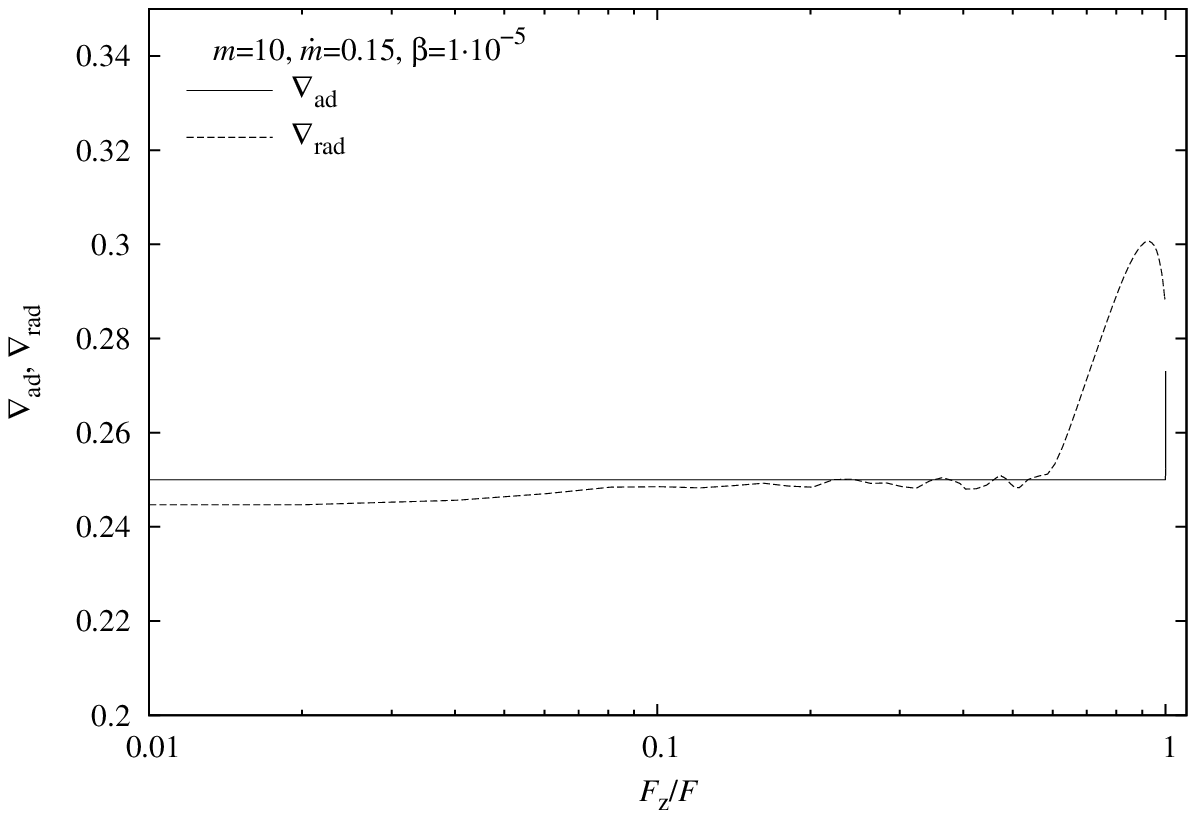}\\%
\vskip-6ex{\small\textbf{(c)}}%
\caption{Vertical disc structure close to the black hole at $s=10r_\textup{\,S}$ for $m=10$, $\dot{m}=0.15$ and $\beta=10^{-5}$}\label{fig_std_z_final_01}
\end{figure}

Nevertheless, we can understand their occurrence by taking a closer look on the vertical structure in the instable zone of the disc.
We therefore plot the vertical stratification of the ratio of the radiation pressure to the total pressure $p_\textup{rad}/p_\textup{tot}$,
the convection efficiency $\zeta$ and the two gradients $\nabla_\textup{rad}$, $\nabla_\textup{ad}$
at a radial position close to the black hole, $s=10r_\textup{S}$ (Fig.~\ref{fig_std_z_final_01}). The data corresponds to the case
$M_\textup{c}=10M_\odot$, $\dot{M}=0.15 \dot{M}_\textup{E}$ and $\beta=10^{-5}$, which showed significant oscillations of $\zeta$ (Fig.~\ref{fig_Mdotvar_final_01}).
For the horizontal axis, we use the heat flux $F_z$ in units of the total flux $F$, given by the energy equation~\eqref{eqn_energy}. The data is taken from a
single solution of the vertical structure without any smoothing or averaging.

For reference, we also display the ratio $p_\textup{rad}/p_\textup{tot}$ at radial positions $100 r_\textup{S}$ and $300r_\textup{S}$.
Close to the black hole, radiation pressure dominates over gas pressure, while they equal each other at $s \gtrapprox 100 r_\textup{S}$.
Further outwards, the disc is gas pressure dominated. The average contribution of convection to the energy transport at these positions
being $\zeta_\textup{avg}=\{0.52,0.1,0\}$, we conclude that convection is radiation pressure driven and by this confirm the results of \citet{shakura_1978}.
In all three cases, the vertical layering of the pressure shows smooth curves. Hence, $\rho$ and $T$ must also adopt such a smooth structure
and the instabilities in $\zeta$ cannot be caused by numerical noise in the density or temperature stratification.

Let us now have a look at the vertical layering of the convective efficiency $\zeta$: we find narrow ``convective cells'' for
small $F_z \approx 0.1 F$-$0.5F$. These small cells are fluctuating for successive iterations (contrary to the extended convective layer between $0.5F$ and $F$),
with the vertical layers close to the mid plane being either fully convective or non-convective.
The reason for these fluctuations can be understood from the lower panel of Fig.~\ref{fig_std_z_final_01}, where we display the two gradients
$\nabla_\textup{rad}$ and $\nabla_\textup{ad}$, which determine whether convection takes place in the disc: the quantity $\zeta$ is
determined by the cubic equation~\eqref{eqn_zeta_disc}, which depends strongly on $B$, with $\zeta \to 0$ for $B \to 0$ and $\zeta \to 1$ for
$B \to \infty$. The key point is that $B$ reflects the Schwarzschild criterium, implying that if the radiative gradient
is less or equal to the adiabatic one, the stratification is stabilized and no convection occurs:
\[
\nabla_\textup{rad} \leq \nabla_\textup{ad} \ \Longrightarrow\ B=0,\qquad \zeta=0\,.
\]
In the opposite case, even a small positive difference $\nabla_\textup{rad}-\nabla_\textup{ad}$ is multiplied by $A^2 \approx 10^{10}$
and therefore $B \gg 1$ and $\zeta \lessapprox 1$. Thus, fluctuations in $\nabla_\textup{rad}-\nabla_\textup{ad}$, regardless of being
of physical or numerical nature, will cause fluctuations in $\zeta$. These fluctuations can not be seen in the physical
quantities, because they occur only for small values of $F_z$ and therefore have little effect on the overall structure. For
a first interpretation of the physics and a relation to observable quantities, these instabilities play only a minor role and
can be replaced by smoothed values. A further investigation of the nature of these fluctuations, however, is necessary in future work.
For instance, we find that the results showing radial variations in $\zeta$ also show an inversion in the vertical layering of the density.
Contrary to the stellar case, it is not clear whether discs are stable when a dominant fraction of the vertical structure is contained
within the inversion region \citep{cannizzo_1988}. 

Furthermore, the assumption of hydrostatic equilibrium becomes questionable as soon as the discs are no longer thin and the vertical motion is
no longer negligible. With a ratio of up to $h/s\approx0.3$ for small supporting $\beta$-viscosities and for small radii $s<100r_\textup{S}$, the
resulting discs should be classified as ``slim'' rather than ``thin''. Interestingly, these cases coincide with the solutions showing strong
radial variations of $\zeta$. Thus, in accordance to \citet{jiao_2008}, waving the assumption of hydrostatic equilibrium and including vertical
motion might be important for a more detailed investigation of these irregularities.
\section{Discussion and conclusion}\label{sec_discussion}
\subparagraph*{Lower limit on the $\beta$-parameter} In the light of the above results, we conclude that convection alone cannot account
for viscosity in accretion discs. It requires an underlying viscosity, produced by some other process, which is parametrized by $\nu_\beta$
in our model. The reason for this can be understood from the following line of argumentation: convection works towards establishing an
adiabatic vertical stratification of the disc. Assuming that there exists an additional source of viscosity in the disc, the convective
elements are decelerated by this inherent friction as well and an equilibrium state is established where energy is transported steadily by
both radiation and convection, and where the total viscosity is given by the sum of the underlying and the convective viscosity. 
If, however, the underlying viscosity is too weak, convection is unchecked and very efficient in building an adiabatic stratification in the
disc with $\nabla_\textup{rad} \lessapprox \nabla_\textup{ad}$. In such a \emph{marginally Schwarzschild-stable} state, no energy is transported
and convection ceases. Thus, convective turbulence and viscosity vanish.

In addition, the total viscosity as a result of the vertical integration over $\nu_\beta+\nu_\textup{conv}$ becomes very small for small values of the supporting $\beta$-viscosity.
This contradicts the requirements from the radial structure equations: the total amount of energy, released by the accretion process and given by~\eqref{eqn_energy}, needs to
be transported away. Furthermore, viscosity must be present to fulfill the angular momentum transport equation~\eqref{eqn_ang_momentum}.
Within this argumentation, the underlying viscosity can also be regarded as the ``driving force'' for convection.

For low supporting $\beta$-viscosities, the density inversion of the vertical layering are more pronounced and at the same time, the disc is
no longer thin. A detailed investigation of the vertical structure is needed for a final conclusion about the lower limit for the supporting
viscosity. In the limits of this investigation, we conclude that an underlying viscosity is necessary and that its minimum value corresponds to
a $\beta$-viscosity with $\beta\approx10^{-5}$. This value agrees well with laboratory measurements of turbulence induced by differential rotation
\citep{richard_1999,richard_2001}.
\subparagraph*{Influence of central mass and accretion rate} Our results show that
the effects of a varying central mass with fixed \emph{absolute} accretion rate are very similar to those of an inversely varying accretion rate
with fixed central mass. With increasing $\dot{M}/\dot{M}_\textup{E}$, the required amount of energy
and angular momentum that has to be transported through the disc increases, leading to larger threshold values for the
total viscosity. Since convection can only partly account for the required increase, the supporting viscosity needs to be larger as well.

Let us consider the case of a varying central mass while the Eddington ratio is kept constant.
Here, changes in $M_\textup{c}$ have a strong influence on the resulting discs, in particular on the density at the disc surface, the effective
temperature and the importance of self-gravity. Figure~\ref{fig_Mcenvar_final_01}d suggests that we can assume that the radial scaling law 
\eqref{eqn_scaling_law_Mdisc} for $M_\textup{d}(s)/M_\textup{c}$ holds for
higher central black hole masses as well. As discussed in Sect.~\ref{sec_central_mass}, the ratio $M_\textup{d}(s)/M_\textup{c}$ also scales with $M_\textup{c}$.
Thus, the estimated disc mass at $s=500 r_\textup{S}$ increases from $10^{-11} M_\textup{c}$ for a stellar mass black hole with $10 M_\odot$ to
$10^{-4} M_\textup{c}$ for a supermassive black hole with $10^8 M_\odot$, boldly assuming that the extrapolation is valid up to this mass.
Correspondingly, the equality radius $s_\textup{equ}$ shrinks by a factor $10^{4}$. As indicated weakly in Fig.~\ref{fig_Mcenvar_final_01}a, higher central masses in principle allow for
lower supporting viscosities due to the lower temperatures and a relatively stronger supporting $\beta$-viscosity (c.\,f., Sect.~\ref{sec_central_mass}).
\subparagraph*{Convective turbulence, differential rotation and magneto-rotational instability: a speculative viscosity-mixture}
Our results reveal that disc solutions do only exist if viscosity is also provided by effects other than convection. Convection itself can
contribute significantly to the total viscosity, but needs a driving force to establish an equilibrium in energy transport in the vertical direction.

Here, we parametrize the supporting viscosity by a permanent $\beta$-viscosity, where the threshold value of the
standard $\beta$-parameter depends (weakly) on the central mass and (strongly) on the accretion rate. For the case of stellar mass
black hole accreting at $10\%$ of the Eddington rate, we find that $\beta\approx 10^{-5}$ is sufficiently large, in agreement with recent
laboratory experiments of rotating Couette-Taylor flows.

In this work, we completely ignore the turbulence created by the MRI. Today being regarded as the primary candidate for the high
viscosity in accretion discs, some aspects still remain to be clarified (c.\,f., Sect.~\ref{sec_introduction}). For example, as detailed in the introduction,
the question whether the viscosity induced by magnetic effects can be translated into an $\alpha$- or $\beta$-type parametrization is still open.
Let us assume for the moment that a parametrization is possible. For example, \citet*{machida_2004} investigated the case of an accretion disc
around a $10 M_\odot$ black hole and found that the corresponding $\alpha$ is not constant, but approximately decreases linearly with radius:
\[
\alpha \propto \exp\left\{\frac{1}{2 s/r_\textup{S}}\right\} - 0.99,\quad \alpha \to 0.01 + \frac{r_\textup{S}}{2s} \quad \mbox{ for } s\gg r_\textup{S}\,.
\]

Their results have to be used carefully since the absolute values in the fitting formula depend strongly on the disc corona -- a high-temperature and
low-density region, put artificially to prevent disc material to evaporate (Machida, priv. comm.). Supposing that the $s^{-1}$ behavior of the MRI
viscosity is roughly valid, we can draw the following picture involving differential rotation, convection, and magnetic turbulence: in the inner disc region,
convection and differential rotation with a corresponding $\beta$-parameter of $\sim 10^{-5}$ alone do not
produce a sufficiently high viscosity for the low central mass and/or high accretion rate case. However, close to the central black hole, the magnetic turbulence
is strong, resulting in a large viscosity due to the magneto-rotational instability. In the intermediate disc region, a weaker MRI effect adds to convection and differential
rotation to account for the required total viscosity. Finally, in the outer disc region, both magnetic effects and convection become negligible,
but differential rotation is sufficient in generating the less demanding values of the total viscosity. An interesting and important investigation therefore would
be to combine these three sources of viscosity and to examine whether the required viscosity can be generated for a large variety of disc parameters.

In this study, we applied the mixing-length theory to describe the convective processes in the accretion disc. Although being applied successfully to stellar and accretion disc
calculations in the past, this theory has several shortcomings like, e.\,g., the neglect of radiative losses and rotation or the unability to derive the anisotropy and the
mixing-length within the model. We completely ignored convection in radial direction, which potentially has significant influence on the resulting disc structure
through its effects on the radial profile of the mass density, for example. In the case of the thin discs considered here, however, the radial heat flux is negligible and radial convection therefore
not important.

In the past, alternative theories for convection in accretion discs have been proposed, although none of them is fully satisfactory. For example,
\citet{cannizzo_1988} investigated the importance of convective turbulence in cataclysmic variables and compared their results for two different models of
convection, the mixing-length theory and a self-consistent theory of convection in accretion discs \citep{cabot_1987a,cabot_1987b}. Their results showed
important differences in the efficiency of convection in generating viscosity. The interesting project of an investigation of alternative theories of convection in our disc model is therefore
left as future work.
\section*{Acknowledgments}
This work was supported by the International Max Planck Research School for Astronomy
and Cosmic Physics at the University of Heidelberg (IMPRS HD), by the Grant-in-Aid for
the 21st Century COE ``Center for Diversity and Universality in Physics'' from the Ministry
of Education, Culture, Sports, Science and Technology (MEXT) of Japan, and by the Japanese
Society for the Promotion of Science (JSPS).

\label{lastpage}

\begin{thebibliography}{88.}%
\addcontentsline{toc}{chapter}{References}
\bibitem[\protect\citeauthoryear{Agol et al.}{2001}]{agol_2001} Agol E., Krolik J., Turner N.J., Stone J.M., ApJ, 558, 543%--552
\bibitem[\protect\citeauthoryear{Artemova et al.}{1996}]{artemova_1996} {Artemova I.V., Bisnovatyi-Kogan G.S., Bj{\"o}rnsson G., Novikov I.D., 1996, ApJ, 456, 119}
\bibitem[\protect\citeauthoryear{Bisnovatyi-Kogan \&\ Blinnikov}{1977}]{bisnovatyi_1977} Bisnovatyi-Kogan G.S.; Blinnikov S.I., 1977, A\&A, 59, 111%--125
\bibitem[\protect\citeauthoryear{Balbus \&\ Hawley}{1991}]{balbus_1991} Balbus S.A., Hawley J.F., 1991, ApJ, 376, 214
\bibitem[\protect\citeauthoryear{Balbus \&\ Hawley}{1998}]{balbus_1998} Balbus S.A., Hawley J.F., 1998, Rev. Mod. Phys., 70, 1
\bibitem[\protect\citeauthoryear{Balbus}{2005}]{balbus_2005} Balbus S.A., 2005, ASPC, 330, 185
\bibitem[\protect\citeauthoryear{Begelman \&\ Pringle}{2007}]{begelman_2007} Begelman M.C., Pringle J.E., 2007, MNRAS, 375, 1070%-1076
\bibitem[\protect\citeauthoryear{Bell \&\ Lin}{1994}]{bell_1994} Bell K.R., Lin D.N.C., 1994, ApJ, 427, 987
\bibitem[\protect\citeauthoryear{B{\"o}hm-Vitense}{1958}]{boehm_1958} B{\"o}hm-Vitense E., 1958, Zs.\,Ap., 46, 108
\bibitem[\protect\citeauthoryear{Brandenburg}{2008}]{brandenburg_2008} Brandenburg A., 2008, Physica Scripta, 130, 014016
\bibitem[\protect\citeauthoryear{Cabot et al.}{1987a}]{cabot_1987a} Cabot W., Canuto V.M., Hubickyj O., Pollack J.B., 1987a, Icarus, 69, 387
\bibitem[\protect\citeauthoryear{Cabot et al.}{1987b}]{cabot_1987b} Cabot W., Canuto V.M., Hubickyj O., Pollack J.B., 1987b, Icarus, 69, 423
\bibitem[\protect\citeauthoryear{Cannizzo \&\ Cameron}{1988}]{cannizzo_1988} Cannizzo J.K., Cameron A.G.W., ApJ,330, 327
\bibitem[\protect\citeauthoryear{Chandrasekhar}{1960}]{chandrasekhar_1960} Chandrasekhar S., 1960, PNAS, 46, 253
\bibitem[\protect\citeauthoryear{Cox \&\ Giuli}{1968}]{cox_1968} Cox J.P., Giuli R.T., 1968, Principles of stellar structure, Vol.~1, Physical Principles, Gordon \&\ Breach,
New-York--London--Paris
\bibitem[\protect\citeauthoryear{Duschl}{1989}]{duschl_1989} Duschl W.J., 1989, A\&A, 225, 105
\bibitem[\protect\citeauthoryear{Duschl, Strittmatter \&\ Biermann} {Duschl et al.}{1998}]{duschl_1998} Duschl W.J., Strittmatter P.A., Biermann P.L., 1998, 192nd AAS Meeting, \#66.17, Bulletin of the American Astronomical Society, Vol. 30, p. 917
\bibitem[\protect\citeauthoryear{Duschl, Strittmatter \&\ Biermann} {Duschl et al.}{2000}]{duschl_2000} Duschl W.J., Strittmatter P.A., Biermann P.L., 2000, A\&A, 357, 1123
\bibitem[\protect\citeauthoryear{Ferguson et al.}{2005}]{ferguson_2005} Ferguson J.W., Alexander D.R., Allard F., Barman T., Bodnarik J.G., Hauschildt P.H., Heffner-Wong A., Tamanai A.,
2005, ApJ, 623, 585
\bibitem[\protect\citeauthoryear{Ferguson}{2008}]{ferguson_www} Ferguson J.W., 2008, Research in Low Temperature Astrophysics at Wichita State University,
\url{http://webs.wichita.edu/physics/opacity/}
\bibitem[\protect\citeauthoryear{Gammie}{1996}]{gammie_1996} Gammie C.F., 1996, ApJ, 457, 355
\bibitem[\protect\citeauthoryear{Goldman \&\ Wandel}{1995}]{goldman_1995} Goldman I., Wandel A., 1995, ApJ, 443, 187
\bibitem[\protect\citeauthoryear{Grevesse \&\ Sauval}{1998}]{grevesse_1998} Grevesse N., Sauval A.J., 1998, Space Science Reviews, 85, 161
\bibitem[\protect\citeauthoryear{Heinzeller}{2008}]{heinzeller_2008} Heinzeller D., 2008, PhD thesis, Univ. Heidelberg, \url{http://www.ub.uni-heidelberg.de/archiv/8575/}
\bibitem[\protect\citeauthoryear{Henyey, Forbes \&\ Gould}{Henyey et al.}{1964}]{henyey_1964} Henyey L.G., Forbes J.E., Gould N.L., 1964, ApJ, 139, 306
\bibitem[\protect\citeauthoryear{Hofmann}{2005}]{hofmann_2005} Hofmann J., 2005, Diploma thesis, Univ. Heidelberg
\bibitem[\protect\citeauthoryear{Jiao et al.}{2008}]{jiao_2008} Jiao C.-L., Xue L., Gu W.-M., Lu J.-F., 2008, \url{http://arxiv.org/abs/0811.2451v1}
\bibitem[\protect\citeauthoryear{King et al.}{2007}]{king_2007} King A.R., Pringle J.E., Livio M., 2007, MNRAS, 376, 1740
\bibitem[\protect\citeauthoryear{Lesur \&\ Longaretti}{2007}]{lesur_2007} Lesur G., Longaretti P.-Y., 2007, in Bouvier J., Chalabaev A., Charbonnel C., eds,
Proceedings of the Annual meeting of the French Society of Astronomy and Astrophysics, Grenoble, France, p. 501
\bibitem[\protect\citeauthoryear{Machida, Nakamura \&\ Matsumoto}{Machida et al.}{2004}]{machida_2004} Machida M., Nakamura K., Matsumoto R., 2004, PASJ, 56, 671
\bibitem[\protect\citeauthoryear{Mineshige \&\ Umemura}{1997}]{mineshige_1997} Mineshige S., Umemura M., 1997, ApJ, 480, 167
\bibitem[\protect\citeauthoryear{Novikov \&\ Thorne}{1973}]{novikov_1973} Novikov I.D., Thorne K.S., 1973, in Witt C.D., Witt B.S.D., eds, Black HolesÑLes Astres Occlus,
Gordon \&\ Breach, New York, p. 343
\bibitem[\protect\citeauthoryear{Paczy\'{n}ski \&\ Wiita}{1980}]{paczynski_1980} Paczy{\'n}ski B., Wiita P.J., 1980, A\&A, {88}, 23%--31
\bibitem[\protect\citeauthoryear{Pessah, Chan \&\ Psaltis}{Pessah et al.}{2007}]{pessah_2007} Pessah M.E., Chan C.-K., Psaltis D., 2007, ApJ, 668 L51%-L54
\bibitem[\protect\citeauthoryear{Pessah, Chan \&\ Psaltis}{Pessah et al.}{2008}]{pessah_2008} Pessah M.E., Chan C.-K., Psaltis D., 2008, MNRAS, 383, 683%-690
\bibitem[\protect\citeauthoryear{Prendergast \&\ Burbidge}{1968}]{prendergast_1968} Prendergast K.H., Burbidge G.R., 1968, ApJ, 151, L83
\bibitem[\protect\citeauthoryear{Pringle \&\ Rees}{1972}]{pringle_1972} Pringle J.E., Rees M.J., 1972, A\&A, 21, 1
\bibitem[\protect\citeauthoryear{Reyes-Ruiz, P{\'e}rez-Tijerina \&\ S{\'a}nchez-Salcedo}{Reyes-Ruiz et al.}{2003}]{reyes-ruiz_2003}
Reyes-Ruiz M. P{\'e}rez-Tijerina E., S{\'a}nchez-Salcedo F.J., 2003, RMxAC, 18, 92R
\bibitem[\protect\citeauthoryear{Richard \&\ Zahn}{1999}]{richard_1999} Richard D., Zahn J.-P., 1999, A\&A, 347, 734%-738
\bibitem[\protect\citeauthoryear{Richard}{2001}]{richard_2001} Richard, D., 2001, Insta\-bi\-li\-t\'{e}s Hydro\-dy\-na\-mi\-ques dans les Ecoule\-ments
en Ro\-ta\-tion Diff\'{e}\-ren\-ti\-elle, PhD thesis, Paris
\bibitem[\protect\citeauthoryear{Ruden et al.}{1988}]{ruden_1988} Ruden S.P., Papaloizou J.C.B., Lin D.N.C., 1988, ApJ, 329, 739
\bibitem[\protect\citeauthoryear{Ryu \&\ Goodman}{1992}]{ryu_1992} Ryu D., Goodman J., 1992, ApJ, 388, 438
\bibitem[\protect\citeauthoryear{Shakura \&\ Sunyaev}{1973}]{shakura_1973} Shakura N.I., Sunyaev R.A., 1973, A\&A, {24}, 337%--355
\bibitem[\protect\citeauthoryear{Shakura, Sunyaev \&\ Zilitinkevich}{Shakura et al.}{1978}]{shakura_1978} Shakura N.I., Sunyaev R.A., Zilitinkevich S.S., 1978, A\&A, 62, 179%--187
\bibitem[\protect\citeauthoryear{Taylor}{1936}]{taylor_1936} Taylor G.I., 1936, Proc. Roy. Soc. London A, 157, 546
\bibitem[\protect\citeauthoryear{TOPS}{2008}]{tops_www} TOPS Astrophysical Opacities: Los Alamos National Laboratory, 2008,, Atomic and Optical Theory,
\url{http://www.t4.lanl.gov/cgi-bin/opacity/astro.pl}
\bibitem[\protect\citeauthoryear{Vehoff}{2005}]{vehoff_2005} Vehoff S., 2005, Diploma thesis, Univ. Heidelberg
\bibitem[\protect\citeauthoryear{Velikhov}{1959}]{velikhov_1959} Velikhov E.P., 1959, J. Exptl. Theoret. Phys., 36, 1398
\bibitem[\protect\citeauthoryear{Vila}{1981}]{vila_1981} Vila S.C., 1981, ApJ, 247, 499
\bibitem[\protect\citeauthoryear{Weiz\-s\"{a}cker}{1948}]{weizsaecker_1948} Weizs\"{a}cker C.F., 1948, Z. Naturforsch, 3a, 524
\bibitem[\protect\citeauthoryear{Wendt}{1933}]{wendt_1933} Wendt F., 1933, Ingenieur-Archiv, 4, 577
\end{thebibliography}
\end{document}